\documentclass[apjl]{emulateapj}
\usepackage{apjfonts}
\usepackage{epsfig}

\def\beq{\begin{equation}}
\def\eeq{\end{equation}}
\def\gauss{{\rm G}}

\begin{document}

\title{Stellar Explosions by Magnetic Towers}
\author{Dmitri A.\ Uzdensky\altaffilmark{1} \& 
Andrew I.\ MacFadyen\altaffilmark{2}}
\altaffiltext{1}{Princeton University, Department of Astrophysical Sciences,
Peyton Hall, Princeton, NJ 08544 --- Center for Magnetic Self-Organization 
(CMSO); {\tt uzdensky@astro.princeton.edu}.}
\altaffiltext{2}{Institute for Advanced Study, Princeton, NJ 08540;
{\tt aim@ias.edu}.}

\date{\today}

\begin{abstract}

We propose a magnetic mechanism for the collimated explosion of massive stars
relevant for long-duration gamma-ray bursts (GRBs), X-ray Flashes (XRFs) and
asymmetric core collapse supernovae.  In particular, we apply Lynden-Bell's
magnetic tower scenario to the interior of a massive rotating star after the
core has collapsed to form a collapsar with a black hole accretion disk or a
millisecond magnetar as the central engine.  The key element of the model is
that the pressure of the toroidal magnetic field, continuously generated by
differential rotation of the central engine, drives a rapid expansion which
becomes vertically collimated after lateral force balance with the
surrounding gas pressure is reached.  The collimation naturally occurs
because hoop stress concentrates magnetic field toward the rotation axis and
inhibits lateral expansion without affecting vertical expansion.  This leads
to the growth of a self-collimated magnetic structure which Lynden-Bell
termed a \emph{magnetic tower}.  When embedded in a massive star, the
supersonic expansion of the tower drives a strong bow shock behind which an
over-pressured cocoon of shocked stellar material forms, as observed in
hydrodynamical simulations of collapsar jets.  The cocoon confines the tower
by supplying collimating pressure support and provides stabilization against
disruption due to magnetohydrodynamical instabilities.  Because the tower
consists of closed field lines starting and ending on the central engine,
mixing of baryons from the cocoon into the tower is suppressed.  Baryon
loading due to magneto-centrifugal winds from the central engine may also be
suppressed because of the expected field line geometry.  The channel cleared
by the growing tower is thus plausibly free of baryons and allows the escape
of magnetic energy from the central engine through the star.  While
propagating down the stellar density gradient, the expansion of the tower
accelerates and becomes relativistic. At some point during the expansion fast
collisionless reconnection becomes possible. The resulting dissipation of
magnetic energy into accelerated particles may be responisble for
GRB prompt emission.
\vspace{0.1 in}
\end{abstract}


%**********************************************************************

\section{Introduction}
\label{sec-intro}

Long duration gamma-ray bursts (GRBs) are asymmetric explosions (Harrison et
al. 1999; Stanek et al. 1999) associated with the death of massive stars
(Matheson et al. 2003; Hjorth et al. 2003; Stanek et al. 2003). Spectroscopic
observations of the afterglow of GRB030329 revealed a bright underlying Type
Ibc supernova, SN2003dh (Matheson et al. 2003; Stanek et al. 2003), which
bore close resemblance to SN1998bw, the first supernova associated with a
GRB, the faint nearby (40 Mpc) GRB980425 (Galama et al. 1998).  Two other
convincing spectroscopic identifications of supernovae associated with
XRF020903 (Soderberg 2005) and GRB031203 (Malesani et al. 2004) have been
reported in addition to about ten photometric detections (Zeh, Klose \&
Hartmann 2004; Soderberg et al. 2006a).  It is now firmly established
observationally that most, perhaps all, of the long duration GRBs are
associated with Type Ibc supernovae.\footnote{However, fewer than 10 percent
of SNe Ibc show evidence of harboring a GRB (Soderberg 2006b).}  Evidence for
strong asymmetry in GRB ejecta is derived from the afterglow light curves
which exhibit, after several days, an achromatic transition in the power law
decay of brightness as a function of time, indicating that the explosion was
beamed into jets with opening angle of $\sim 10$ degrees.  Taken together,
these observations show that some supernova explosions involve strongly
asymmetric relativistic outflow.  Massive stars are thus observed to be
capable, in some cases, of producing strongly collimated energetic outflows
when they die.

Such asymmetry is naturally expected in models where the progenitor star 
is rapidly rotating when it collapses and subsequently explodes.  Stellar
rotation breaks spherical symmetry and provides the rotation axis as a
preferred direction along which jet-like outflow can develop.  Observations
of massive stars do indicate rapid surface rotation velocities (Fukuda 1982),
though the distribution of angular momentum in the cores of evolved stars
when they undergo core collapse is uncertain (Spruit 2002; Heger, Woosley \&
Spruit 2005; Maeder \& Meynet 2005).  However, low metallicity is expected 
to be beneficial for maintaining angular momentum in the core until collapse
(MacFadyen \& Woosley 1999; Hirschi, Meynet, \& Maeder~2005).  Current
stellar evolution models do, in some cases, predict rapid rotation in
collapsing cores of massive single stars (Heger, Woosley \& Spruit 2005),
perhaps aided by global mixing (Woosley \& Heger 2005; Yoon \& Langer 2005).
It is also expected that membership in a tight binary system will endow some
massive stars with rapid rotation in their cores when they die (Fryer \&
Heger 2005; Petrovic~et~al.~2005) though the angular momentum imparted to 
the core depends on detailed transport processes (Petrovic~et~al.~2005).

A theoretical model for long duration GRBs based on rapid stellar rotation 
is known as the {\it collapsar} (Woosley~1993; Paczynski 1998; MacFadyen \&
Woosley~1999; MacFadyen, Woosley \& Heger~2001) in which the core of a
massive rotating star collapses to form a black hole.  Numerical simulations
have shown that, in rapidly rotating stars, an accretion disk rapidly forms
around the young black hole (MacFadyen \& Woosley~1999).  Because of angular
velocity gradients in the disk, the magneto-rotational instability (MRI) is
expected to develop, providing angular momentum transport and dissipation of
gravitational energy.  At the temperatures and densities ($T \sim 4 \times
10^{10}$ K, $\rho \sim 10^{10}$ g cm$^{-3}$) present in collapsar disks,
neutrino emission can cool the accreting gas with a range of efficiency. The
collapsing outer stellar core is therefore able to accrete at rates of $\sim
0.1 M_{\odot} s^{-1}$ for times $\gtrsim 10$ s providing sufficient energy
for sufficiently long times to power long GRBs.

Magnetohydrodynamical (MHD) simulations have demonstrated the
development of MRI in collapsar disks (Proga~et~al. 2003; McKinney~2005).  
Magnetic fields approaching equipartition values in excess 
of~$10^{15}$~G are then expected to develop as the disk forms.  
In addition, field strengths approaching $10^{15}$~G may develop 
during collapse as seed poloidal field winds up into toroidal field.  
Similar processes, resulting in similar field strengths, are also 
believed to be taking place in core-collapse supernovae explosions 
(Akiyama~et~al.~2003; Ardeljan, Bisnovatyi-Kogan, \& Moiseenko~2005).

In addition, it is possible that a turbulent dynamo is not required in
the collapsar case. This is because a plausible source of large-scale
poloidal field threading the disk may be the field advected and
compressed from the pre-collapse star. Magnetized white dwarfs are
observed to have surface magnetic fields of~$10^9$~G. If the cores of
some massive cores posses fields of similar magnitude, they will grow
to~$10^{14}$~G if a core of radius~$10^9$~cm and $B\sim 10^9$~G
collapses to $3\cdot 10^6$~cm.  As was recently established
by Braithwaite \& Spruit (2005), stable large-scale poloidal 
field structures can be maintained for a stellar lifetime by
a spheromak-like field configuration involving twisted toroidal 
field inside the star.

Because gas along the rotation axis of the star does not experience a
centrifugal barrier, it falls into the black hole unimpeded, thus lowering
the density above the hole to $\rho\lesssim 10^6$~gm~cm$^{-3}$. Similarly to
its role in Type~II supernovae explosions (Goodman~et~al. 1987), neutrino
annihilation may be important in the low density funnel above the black hole,
heating the stellar gas and causing it to expand along the rotation axis of
the star.  Under favorable conditions this hot gas can form a collimated
relativistic outflow which is capable of escaping the stellar surface and
forming a fireball. MacFadyen \& Woosley (1999) explored this possibility but
noted that magnetic processes are an alternate mechanism capable of
extracting energy from the system.  In fact, the suggestion that strongly
magnetized jets play an active role in GRB explosions (perhaps via a version
of the Blandford--Znajek mechanism) has been invoked by several authors
(e.g., Thompson~1994; Meszaros \& Rees 1997; Lee~et~al. 2000; Vlahakis \&
K{\"o}nigl 2001; van~Putten \& Ostriker 2001; Drenkhahn \& Spruit 2002;
Lyutikov \& Blandford 2003; van~Putten \& Levinson 2003; Proga~et~al. 2003; 
Lyutikov 2004, 2006; Lei~et~al. 2005; Proga \& Zhang 2006).
In addition, a lot of relevant work has been done in the supernova 
context, where several magnetic explosion mechanisms have been explored
(e.g., Leblanc \& Wilson 1970; Meier~et~al. 1976; Wheeler~et~al. 2000, 2002; 
Akiyama~et~al. 2003; Ardeljan~et~al. 2005; Blackman~et~al. 2006).

In this paper, motivated by the above considerations, we consider a possible
magnetic mechanism for driving a low-baryon-load, Poynting-flux dominated jet
through a massive collapsing star. In order to understand the structure of
magnetic jets in massive stars we investigate a simple analytic model that
may help guide interpretation of detailed MHD simulations.

Our paper is organized as follows. In~\S~\ref{sec-magnetic-tower} we describe
the key physical ideas that underlie our model. The main idea is that we
apply the {\it magnetic tower} mechanism, developed previously by Lynden-Bell
(1996) for active galactic nucleus (AGN) jets, to the collapsar
environment. We first give a qualitative description of Lynden-Bell's
original model (in an accretion disk geometry) and derive some basic scalings
resulting from it (\S~\ref{subsec-LB-model}).  In the rest
of~\S~\ref{sec-magnetic-tower}, we present our picture of a magnetic tower
growing rapidly inside a star and driving a strong shock through it. Thus,
in~\S~\ref{subsec-cocoon-shock}, we describe our modifications to the
original magnetic tower model; the most important of these is that the
external pressure confining the tower is no longer arbitrary but is
determined self-consistently by the hot gas {\it cocoon} surrounding the
tower and by the strong shock driven into the star. In~\S~\ref
{subsec-estimates} we derive simple scalings for the tower parameters and
then, in~\S~\ref{subsec-numbers}, we use these scalings to make some
quantitative estimates. In~\S~\ref{sec-equations} we make the next step and
present mathematical formalism of our model.  Our analysis is characterized
by representing the time evolution of the tower by a sequence of force-free
magnetostatic equilibria, obtained by solving the Grad--Shafranov equation
with additional constraints and boundary conditions that, in general, change
with time.  We then present, as examples, several particular analytical
solutions (in~\S~\ref{sec-examples}).  After that,
in~\S~\ref{sec-discussion}, we discuss the implications of our model for
gamma-ray bursts and outline some important open issues that we feel need to
be addressed in future research.  In particular, in~\S~\ref{subsec-rel} we
argue that, as the tower expands in a stratified star, the unperturbed
stellar density at the top of the tower drops and the expansion process
accelerates.  At a certain point, the tower growth will then have to make a
transition to the relativistic regime, which is not covered by our theory.
Nevertheless, we speculate that the tower will remain collimated, with
an opening angle of a few degrees.  In~\S~\ref{subsec-structure} we discuss
the tower structure and the distribution of magnetic energy inside the tower.
In~\S~\ref{subsec-baryons} we discuss the issue of baryon contamination
originating from the disk as a result of neutrino ablation and
magneto-centrifugal winds.  Next, in~\S~\ref{subsec-stability}, we discuss
various MHD instabilities that may be excited in the system and the effect
they may have on the evolution of the magnetic tower.
In~\S~\ref{subsec-reconnection}, we discuss the question of whether the tower
can be disrupted through reconnection while it is still deep inside the star
and argue against this possibility. In~\S~\ref{subsec-numerical} we outline
numerical simulations which may be performed to explore the model. 
We present our conclusions in~\S~\ref{sec-conclusions}.

Finally, note that for definiteness we develop our model in the context of
the original collapsar scenario, where the central engine is an accretion
disk around a black hole. The energy source in this case is accretion
energy. However, we believe that a similar magnetic tower mechanism may also
work when the central engine is a young millisecond magnetar born inside a
collapsing star. The typical values of the magnetic field strength, the
rotation rate, and the size of the system are similar in the two cases. The
overall electro-magnetic luminosities should therefore also be comparable
(e.g., Usov~1992; Thompson~1994). The energy source for the explosion in the
magnetar case is the rotational energy of the neutron star and the magnetic
field acts as an agent that extracts this energy.  The basic mechanism is
essentially similar to that proposed by Ostriker \& Gunn (1971) for powering
supernovae light-curves by the spin-down magnetic luminosity of a
rapidly-rotating pulsar (with the magnetic field scaled up by three orders of
magnitude and the timescale scaled down by six).  In fact, several models
invoking rapidly-rotating magnetars as central engines for GRBs have been
proposed (e.g., Usov~1992; Thompson~1994; Yi \& Blackman 1998;
Thompson~et~al. 2004; Lyutikov~2006). However, these models usually don't
discuss the geometry (e.g., collimation) of the outflow. In addition, they
typically consider an isolated magnetar (without a surrounding stellar
envelope). The collimating influence of the dense stellar material is thereby
neglected.  We believe that a rapidly-rotating magnetar will make an even
better central engine when considered within the collapsar framework. A
simplified physical model representing this scenario is the {\it
pulsar-in-a-cavity} problem. By analyzing this problem, we can show that,
even though the magnetar itself rotates uniformly, a strong differential
rotation is effectively established on the field lines extending beyond the
light cylinder. As a result, toroidal magnetic flux is constantly injected
into the cavity and eventually toroidal magnetic field becomes dominant over
both the poloidal magnetic field and the electric field in the
magnetosphere. Any subsequent expansion of the cavity is then going to be
mostly vertical because of the collimating hoop stress, just as in
Lynden-Bell's (1996) model. Eventually, a magnetic tower forms.  In our
opinion, this pulsar-in-a-cavity problem, viewed as a paradigm for a
millisecond-magnetar-driven GRB explosion mechanism, is very important by
itself. Therefore we believe it deserves a separate study and we intend to
pursue it in an upcoming paper.

%**********************************************************************

\section{Magnetic Tower: Basic Physical Picture}
\label{sec-magnetic-tower}

\subsection{Lynden-Bell's Original Magnetic Tower Model}
\label{subsec-LB-model}

\begin{figure}
%\plotone{lb-tower.eps}
\plotone{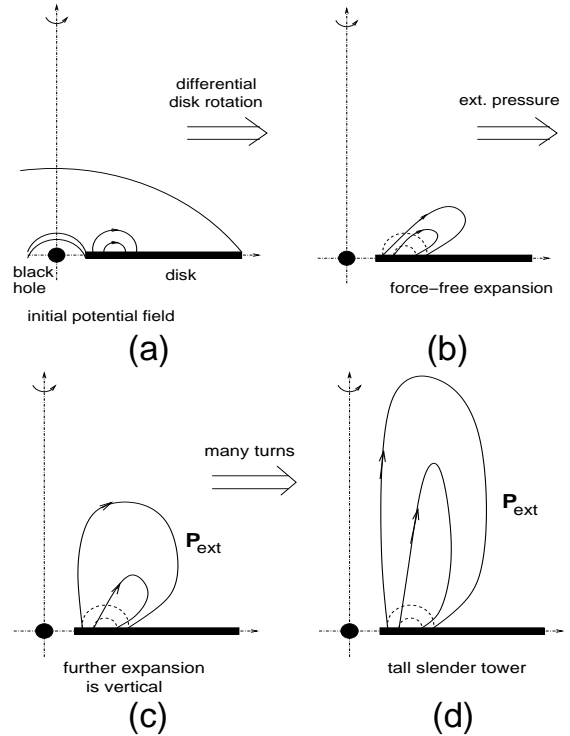}
\figcaption{Development of a magnetic tower in Lynden-Bell's (1996) model.
\label{fig-LB-tower}}
\end{figure}

As a specific mathematical vehicle to illustrate our ideas, we choose the
{\it magnetic tower} introduced by Lynden-Bell (1996, 2003).  A magnetic
tower is an axisymmetric magnetic configuration that arises when a system of
nested closed flux surfaces, anchored in a differentially-rotating disk, is
twisted and, as a result, inflates, but when this inflation is controlled by
a surrounding external pressure.  The basic physical mechanism of this
process can be described as follows (see Fig.~\ref{fig-LB-tower}).

Consider a thin conducting disk with some vertical magnetic flux frozen into
it. Let us assume that initially the magnetic field is potential and has a
fully-closed dipole-like topology (see Fig.~\ref{fig-LB-tower}a), with the
two footpoints of each field line located at different radii on the disk. Now
let us assume that the disk rotates non-uniformly, which is certainly the
case for a Keplerian disk. Then, each field line marked by the 
parameter~$\Psi$ is twisted at a rate
$\Delta\Omega(\Psi)$ equal to the difference in angular velocities of the two
footpoints of this line.  Correspondingly, toroidal magnetic flux is
generated from the poloidal flux. The pressure of this toroidal field pushes
the flux surfaces outward, against the poloidal field tension.  It is assumed
that during the initial stages of this process the gas pressure, as well as
the gravitational and inertial forces, are negligibly small in the disk
magnetosphere, so the magnetic field is force-free. Then the expansion is
uncollimated, typically along the direction making a $60^\circ$ angle with
respect to the rotation axis (Fig.~\ref{fig-LB-tower}b). However, as was
shown by Lynden-Bell (1996), if there is some, no matter how small, external
gas pressure that surrounds the expanding disk magnetosphere, then at some
point the sideways expansion is stopped. This is because the magnetic field
strength at the outermost portions of inflating field lines decreases rapidly
during the inflation process; then, once $B^2/8\pi$ drops to the level of the
external gas pressure, any further horizontal expansion ceases. However, as
again was shown by Lynden-Bell (1996), the story doesn't end here; unable to
expand sideways, the twisted magnetosphere expands in the vertical direction
(Fig.~\ref{fig-LB-tower}c), eventually forming a slender cylindrical column
that Lynden-Bell termed a {\it magnetic tower} (see Fig.~\ref{fig-LB-tower}d). If
the external pressure outside of the tower is kept constant and uniform, then
the top of the tower rises at a constant speed. Plasma inertia never plays
any role in this process; the entire evolution is viewed as a sequence of
magnetostatic equilibria, with the field being force-free inside the tower
and with pressure balance between the external gas outside of the tower and
the magnetic field inside.

Note that the assumption that both ends of a field line connect to the disk
itself is actually not essential. An alternative configuration, with a very
similar overall behavior, is that of a rotating conducting disk magnetically
connected to a highly-conducting rotating central star (a proto-neutron star
in the collapsar context) or even a rotating black hole. In the conducting
star case, the differential star--disk rotation leads to the same field-line
inflation and opening process in the force-free regime (e.g., Lovelace~et~al.
1995; Uzdensky~et~al.~2002), followed by the tower stage when the external
pressure becomes important. A very similar process takes place when a
conducting disk is connected magnetically to a rapidly-rotating black hole:
even though a black hole cannot be considered a good conductor,
general-relativistic frame-dragging ensures that a closed-field configuration
that extends far enough on the disk cannot stay in equilibrium and therefore
has to inflate and open up (Uzdensky~2005). This picture is consistent with 
that observed in recent general-relativistic MHD (GRMHD) numerical simulations 
of accreting black holes (McKinney~2005; Hawley \& Krolik~2006).
 
In order to get accurate quantitative results, one needs to consider a
specific solution of the governing MHD equations with some particular
boundary conditions. However, to get a basic physical feeling of how the
magnetic tower grows, it is instructive to derive some simple
order-of-magnitude estimates and scaling relationships.  In Lynden-Bell's
(1996) model, the main input parameters that determine the solution are the
total poloidal magnetic flux~$\Psi_0$ (per unit toroidal angle) in the tower,
the characteristic differential rotation rate~$\Delta\Omega$, and the
external pressure~$P_{\rm ext}$. In the rest of this section we investigate
how the main parameters that characterize the tower, namely, its radius, the
typical magnetic field, and the growth velocity, scale with these three input
parameters.

First, we estimate the radius of the tower, $R_0$, and the characteristic 
poloidal magnetic field strength, $B_{\rm pol}$. They are related via 
\beq
B_{\rm pol} \sim B_0 \equiv {\Psi_0\over R_0^2} \, .
\label{eq-def-B0}
\eeq
As we mentioned earlier, the radius adjusts so that the magnetic pressure 
inside the tower equals the external gas pressure. From the force-free 
balance inside the tower we expect the toroidal magnetic field to be roughly 
the same as the poloidal field: $B_\phi\sim B_{\rm pol}\sim B_0$. The total 
magnetic field strength at the outer edge of the tower is thus also of the 
order of~$B_0$. Then, from the the condition of pressure balance across the 
tower's side wall we get
\beq
B_0 \sim \sqrt{8\pi P_{\rm ext}} \, . 
\label{eq-LB-side-equil}
\eeq
By combining this with equation~(\ref{eq-def-B0}), we get
\beq
R_0 \sim \biggl( {\Psi_0^2\over{8\pi P_{\rm ext}}}\biggr)^{1\over4} \, .
\label{eq-LB-R0}
\eeq

Next, let us estimate the rate of growth of the tower.  To do this, we use
the fact that the toroidal magnetic flux is continuously generated out of the
poloidal flux by the differential rotation. Every time a given flux tube with
poloidal flux $\Delta\Psi$ is twisted by one full turn, the amount of
toroidal flux carried by the tube is increased by $\Delta\Psi$. Therefore,
after $N=\Delta\Omega t/(2\pi)$ turns the total toroidal flux $\chi$ in the
tower becomes

\beq
\chi = 2\pi \Psi_0 N = \Psi_0 \Delta\Omega t   \, .
\eeq

Assuming that this toroidal flux fills the tower uniformly and that the tower
is a cylinder with radius $R_0$ and height~$Z_{\rm top}$, we see that the
characteristic toroidal field has to be of the order of

\beq
B_\phi \sim {\chi\over{R_0 Z_{\rm top}}} = 
{\Psi_0\over{R_0 Z_{\rm top}}}\, \Delta\Omega t = 
B_0\, {{R_0}\over{Z_{\rm top}}}\, \Delta\Omega t    \, .
\eeq

However, as we stated earlier, the typical toroidal field in the tower is of
the order~$B_0$; therefore, the height of the tower has to increase steadily
as
\beq
Z_{\rm top}(t) \sim R_0\Delta\Omega t \, .
\eeq
In other words, the tower grows at the speed comparable to the typical
differential rotation velocity $R_0\Delta\Omega$. If the external 
pressure does not change, the radius of the tower, determined by
equation~(\ref{eq-LB-R0}), stays constant during its growth; therefore, 
after many turns ($\Delta\Omega t\gg 1$), the height $Z_{\rm top}$ becomes 
much larger than the radius, i.e., the tower becomes slender.

Since the first analytical solution proposed by Lynden-Bell (1996), 
the magnetic tower concept is becoming more and more accepted by the
astrophysical community. For example, the formation and evolution of 
magnetic towers have been studied in numerical simulations (Li~et~al. 2001;
Kato~et~al. 2004) and even in real laboratory experiments (Hsu \& Bellan
2002; Lebedev~et~al. 2005). When comparing with the recent GRMHD simulations 
of accreting black holes (e.g., McKinney 2005; Hawley \& Krolik 2006), one 
should note that there is no confining external gas pressure in these 
simulations, and so one may not expect outflow collimation by the magnetic 
tower mechanism. However, one may expect a poorly collimated tower-like 
magnetic structure on very large scales, with a narrow Blandford--Znajek 
(1977) jet consitituting the inner core of the tower.

%---------------------------------------------------------------------

\subsubsection{Energy Flow in Magnetic Towers}
\label{subsubsec-energy-flow}

An interesting question is how energy flows through a magnetic tower. 
As one can easily see, Poynting flux flows up from the disk along the 
inner segment of each field line and down to the disk along the outer 
segments.
In our view, this can be understood as follows. For each 
field line~$\Psi$, the inner, faster-rotating footpoint~($1$)
performs work on the magnetic field at a rate proportional to 
$W_1\sim I_{\rm pol}[\Psi(1)] \Omega(1)$ (per unit of poloidal 
flux). The corresponding decelerating torque per unit flux is
proportional to $\tau_1\sim  I_{\rm pol}[\Psi(1)]$. In turn, 
the magnetic field exerts an accelerating torque per unit flux
$\tau_2\sim I_{\rm pol}[\Psi(2)]$ on the outer disk footpoint of the same 
field line. Correspondingly, it performs work at a rate $W_2\sim 
I_{\rm pol}[\Psi(2)]\Omega(2)$. Because of force-free equilibrium
in the tower, $I_{\rm pol}[\Psi(1)]=I_{\rm pol}[\Psi(2)]=I_{\rm pol}(\Psi)$, 
and so $\tau_1=\tau_2$, i.e., all the angular momentum extracted magnetically 
from point~$1$ is transferred to point~$2$. The two energy flows, 
on the other hand, are not equal. Indeed, since $\Omega(1)>\Omega(2)$, 
the energy that is extracted from point~$1$ and that flows up along 
the inner segment of the field line is greater than the energy that 
flows down along the outer segment and is deposited in the disk at 
point~$2$. The difference, proportional to $I_{\rm pol}(\Psi)$, is
the actual power that drives the expansion of the tower. A part of 
it goes into filling the growing volume of the tower with magnetic
energy, and the rest goes into performing work against external gas
pressure and driving the shock through the star. 

It is interesting to note that the total vertical Poynting flux in 
the two segments only involves the differential rotation~$\Delta\Omega
=\Omega(1)-\Omega(2)$, but is independent of the absolute rotation itself.
This is because we are dealing here with a force-free equilibrium, so that 
$I_{\rm pol}$ is constant along the entire length of a field line; in 
particular, it has the same sign on the two segments of the field line,
and hence so does~$B_\phi$. The situation would be drastically different
in the relativistic-rotation case, where both field-line segments extend
beyond their respective light cylinders. In that case, one would no longer
have a force-free equilibrium along the entire stretch of a field line;
in particular, equilibrium will most likely break down at the farthermost 
tip of the line where the two segments join. As a result, the signs of
$I_{\rm pol}$ (and hence of~$B_\phi$) on the two segments would be
opposite, which corresponds to both segments being swept back.
Consequently, the Poynting flux would be outward along both segments.
A similar situation arises in the non-force-free MHD case;
the two field-line segments would then be swept back by plasma
inertia if they extend beyond the Alfv\'en point. This would 
again result in an outward Poynting flux along both segments.
In both of these cases, the total vertical Poynting flux depends
on the absolute rotation rates~$\Omega(1)$ and~$\Omega(2)$ themselves, 
as opposed to just their difference.

%***************************************************************

\subsection{Magnetic Tower Driving a Shock through a Star}
\label{subsec-cocoon-shock}

As we mentioned briefly in the Introduction, we believe that there are
several good reasons that make the magnetic tower mechanism an
attractive model for the formation and propagation of a
magnetically-dominated jet through a star within the collapsar model
for GRBs (and perhaps core-collapse supernovae as well). To reiterate,
a configuration where all the field lines close back onto the central
engine (e.g., an accretion disk or a magnetar) is very natural for a
field created by a turbulent dynamo (with zero net flux). In addition,
diffusion of particles across the field is suppressed and thus baryon
contamination of the tower from the stellar envelope is inhibited.
This is in contrast with the models that invoke the Blandford--Znajek
mechanism operating directly along open field lines inside the star.
Even if a light, ultra-relativistic outflow is launched along such
field lines, it will first have to push all the baryonic matter,
already present on these field lines, ahead of itself; therefore, the
resulting ejecta in these models will not be baryon-free. On the other
hand, the closed-field geometry, characteristic of the magnetic tower
model, at least eliminates this problem. The only way baryons can get
into the tower is via a wind blowing from the disk or the neutron
star.  An assessment of baryon contamination due to a wind is an
important issue that we will discuss in~\S~\ref{subsec-baryons}.

\begin{figure}
%\plotone{star.eps}
\plotone{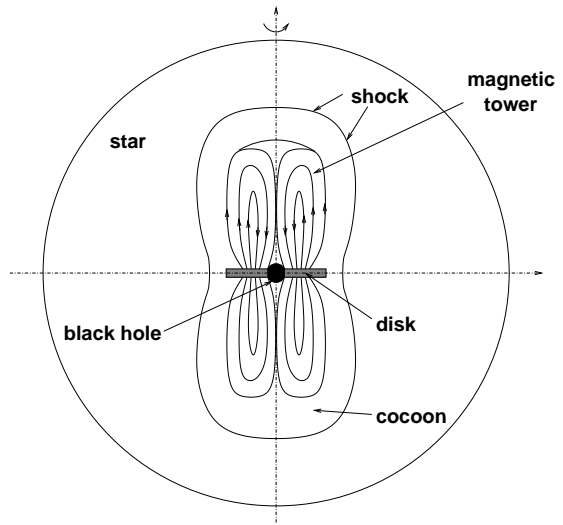}
\figcaption{The main components of our model. The lines attached to the disk
represent poloidal magnetic field lines (toroidal magnetic field is also
present). A magnetic tower grows rapidly and drives a strong shock through 
the star. The shocked stellar gas behind the shock forms a hot cocoon whose 
high pressure confines the tower.
\label{fig-star}}
\end{figure}

In order to apply the magnetic tower model to the collapsar scenario,
we first have to make a few modifications to Lynden-Bell's original
picture. Specifically, we extend his model by taking into account the
high-pressure cocoon that surrounds and confines the tower.  In our
model, the magnetic tower grows very rapidly and thus acts as a piston
driving a shock ahead of it. The shocked stellar gas above the tower
has very high pressure; it therefore squirts sideways and forms
backflows that fill the cocoon around the tower as described for AGN
jets by Begelman, Blandford \& Rees (1984) and in the collapsar case
by Ramirez-Ruiz, Celotti \& Rees (2002) and Matzner (2003) (see
Fig.~\ref{fig-star}).  Therefore, the external pressure that confines
the tower is no longer an arbitrary parameter, in contrast with
Lynden-Bell's model. Instead, it is determined by the jump conditions
across the shock surrounding the cocoon across the contact
discontinuity between the cocoon and the tower.  Since the speed at
which the magnetic tower plows through the star is much higher than
the sound speed in the unperturbed stellar material, the external
unperturbed pressure of the star is irrelevant; it should thus be
excluded from our list of the three main input parameters used for our
simple, order-of-magnitude estimates.  Instead, the expansion of the
cocoon is controlled by the ram pressure related to the gas {\it
inertia}; therefore, we replace the external pressure by the
unperturbed stellar density~$\rho_0$ in the list of basic dimensional
parameters (along with $\Psi_0$ and~$\Delta\Omega$) that determine the
relevant physical scales in our problem. This change represents an
important difference between our model and Lynden-Bell's.

The actual situation is complicated further by the two-dimensional 
character of the problem. Indeed, since the sound speed inside the 
cocoon is very high, comparable to the speed at which the tower grows, 
gas pressure tends to be partially equalized throughout the cocoon. 
This expectation is supported by the hydrodynamic numerical simulations%
\footnote
{Note that, even though the simulations by MacFadyen~et~al. (2001) and
Zhang~et~al. (2003) do not include magnetic fields, they probably
still give a reasonably good qualitative representation of the cocoon
behavior even in our magnetic tower scenario, since the cocoon itself
remains unmagnetized.}  of the collapsar model by MacFadyen~et~al.
(2001) and Zhang~et~al. (2003), which show that, typically, the
variation of the cocoon pressure along its length is relatively weak,
just a factor of~5 or~10. This is very moderate compared to the
corresponding variations of the unperturbed stellar density and
pressure, which both vary by many orders of magnitude along the
vertical extent of the cocoon.  As a result, the gas pressure is very
high everywhere in the cocoon so the cocoon also drives a sideways
shock into the star.  Thus, the boundary between the tower and the
cocoon is a contact discontinuity, whereas the boundary between the
cocoon and the rest of the star is a two-dimensional strong shock of
some complicated shape.

We see that, in principle, the problem calls for a consideration of 
the entire {\it two-dimensional} shock-cocoon-tower structure. For simplicity, 
however, we shall represent the cocoon by only two components: 
the hot spot right above the tower (with some uniform pressure~$P_{\rm top}$) 
and the cylindrical shell surrounding the tower on the sides (with a 
different uniform pressure~$P_{\rm side}$). Thus, we shall parameterize 
the pressure non-uniformity in the cocoon by the ratio 
$\eta=P_{\rm side}/P_{\rm top}$. 
In accordance with this 2-component cocoon structure, we shall represent
the strong shock between the cocoon and the star also by two components:
a plane shock that propagates purely vertically above the 
tower and a slower cylindrical shock that propagates sideways. An important 
difference between these two shocks is that, whereas the upward shock is 
driven exclusively by the vertical growth of the magnetic tower, the 
sideways shock is driven mostly by the pressure of the hot gas that 
flows into the side part of the cocoon from the top hot spot;
the horizontal expansion of the tower may also play a role in
driving the sideways shock, but probably to a lesser extent.

Let us now discuss what determines~$\eta$. We use simple energetics
considerations for a rough estimate. The energy flux that is required to
drive the sideways shock is of order $P_{\rm side} V_{\rm s,side}$, where
$V_{\rm s,side}$ is the speed of the sideways shock; it is, in turn,
proportional to the sound speed, $V_{\rm s,side}\sim(P_{\rm side}/\rho_{0,\rm
side})^{(1/2)}$.  In our simple 2-component model, the area of the sideways
shock is larger than the area of the vertical shock by a factor $\kappa\equiv
Z_{\rm top} R_{\rm cocoon}/R_0^2\gg 1$.  Thus, in order for the power
necessary to maintain the sideways shock not to exceed the available thermal
power generated by the vertical growth of the tower, one must satisfy
\beq
\eta \leq \kappa^{-2/3}\, 
\biggl({\rho_{0,\rm side}\over{\rho_{0,\rm top}}}\biggr)^{1/3} \, .
\label{eq-eta}
\eeq
Thus, we see that, as the tower grows (and~$\kappa$ increases) with time, 
$\eta=P_{\rm side}/P_{\rm top}$ tends to decrease, but, on the other hand, 
this decrease is partially weakened by the density non-uniformity, i.e., 
by the large ratio~$\rho_{0,\rm side}/\rho_{0,\rm top}$.
We also see that the sideways shock propagates noticeably slower 
than the vertical shock. This is due to a combination of two 
factors: a smaller (by a factor of~$\eta$) pressure driving the sideways  shock
and a higher background density of the stellar gas into which the sideways shock 
is propagating. As a result, the cocoon is moderately elongated in the 
vertical direction, with an aspect ratio of 
\beq
{Z_{\rm cocoon}\over{R_{\rm cocoon}}} \sim
{V_{s,\rm top}\over{V_{s,\rm side}}} \sim
\eta^{-1/2}\, \sqrt{\rho_{0,\rm side}\over{\rho_{0,\rm top}}} \sim
\biggl(\kappa\,{\rho_{0,\rm side}\over{\rho_{0,\rm top}}}\biggr)^{1/3}\, .
\label{eq-cocoon-aspect-ratio}
\eeq

These estimates, of course, are only as good as the underlying
two-component cocoon structure assumed here. In a realistic situation,
where the tower is growing in a strongly stratified medium, the radii 
of both the tower and the cocoon are not going to be constant in the vertical 
direction. In that situation, one cannot represent the tower and the 
cocoon by straight cylinders; instead, one has to describe their shapes
by some increasing functions~$R_0(z)$ and~$R_{\rm cocoon}(z)$.

%*****************************************************************

\subsection{Simple Estimates}
\label{subsec-estimates}

Let us now show how the basic properties of the growing magnetic tower 
are expressed in terms of the three parameters~$\Psi$, $\Delta\Omega$,
and~$\rho_0$. For simplicity, in this and the next subsections, we ignore 
the non-uniformity of the gas pressure in the cocoon, i.e., we set~$\eta=1$. 
We will reinstate this non-uniformity and study the dependence of the tower 
growth parameters on~$\eta$ in our more rigorous analysis of~\S\S~\ref
{sec-equations}--\ref{sec-examples}. 

For simplicity we shall use shock jump conditions corresponding to a 
one-dimensional problem. Since the gas pressure of the unperturbed 
stellar material upstream of the shock is neglected, the shock is 
strong. Assuming a gas with an adiabatic index~5/3, 
the shock velocity with respect to the unperturbed gas is $V_s=4/3 \,
V_{\rm top}$, whereas the pressure in the post-shock region (i.e., in 
the cocoon) can be expressed in terms of the velocity of the piston 
$V_p\equiv V_{\rm top}$ and the upstream gas density~$\rho_0$ as (see, 
e.g., Kulsrud~2005) 
\beq
P_{\rm top} = {3\over 4}\, \rho_0 V_s^2 = {4\over 3}\,\rho_0 V_{\rm top}^2 \, .
\label{eq-shock-pressure}
\eeq

By comparing this result with the condition of pressure balance 
$P_{\rm top}\simeq B_0^2/8\pi$ at the contact discontinuity at the top 
of the tower, we immediately see that the tower grows with a velocity 
of order the Alfv\'en speed computed with the unperturbed density~$\rho_0$:
\beq
V_{\rm top} \sim V_A \equiv {B_0\over\sqrt{4\pi\rho_0}} =
{\Psi_0\over{R_0^2\sqrt{4\pi\rho}}}   \, .
\label{eq-V_top=V_A}
\eeq

On the other hand, as we have shown earlier, $V_{\rm top}$
should be of the order of the differential rotation speed~$R_0\Delta\Omega$.
Thus, we obtain the following estimate for the radius of the tower
in terms of~$\Psi_0$, $\Delta\Omega$, and~$\rho_0$:
\beq
R_0 \sim 
\biggl({\Psi_0\over{\Delta\Omega}}\biggr)^{1/3}\, (4\pi\rho_0)^{-1/6} \, .
\label{eq-R0}
\eeq
We can also relate the radius of the tower to the radius of the central 
accretion disk. The poloidal flux~$\Psi_0$ can roughly be estimated as 
\beq
\Psi_0 \sim B_d R_d^2      \, ,
\label{eq-Psi0-Bd}
\eeq
where $B_d$ is the typical magnetic field in the disk 
(or, rather, its dipole-like part), and $R_d$ is the
characteristic radius of the inner part of the disk
--- the base of the tower. Then,
\beq
B_0 \sim B_d\, {{R_d^2}\over{R_0^2}}  \, ,
\label{eq-B0-Bd}
\eeq
and 
\beq
{R_0\over{R_d}} \sim \biggl({{\tilde{V}_{\rm A,d}}\over{V_d}}\biggr)^{1/3} \, ,
\label{eq-R0-Rd}
\eeq
where $V_d \equiv R_d \Delta\Omega$ is the characteristic differential 
rotation velocity in the disk (on the order of the Keplerian velocity~$V_K$) 
and 
\beq
\tilde{V}_{\rm A,d} \equiv {B_d\over{\sqrt{4\pi\rho_0}}} 
\eeq
is a convenient composite quantity that has a form of an Alfv\'en 
speed involving the disk magnetic field and the unperturbed star's
plasma density; it doesn't represent a physical velocity and so
can be arbitrarily high; in particular, it can be greater than 
the speed of light.

Notice that as the tower makes its way through the star, 
two of the three main parameters, $\Psi_0$, and $\Delta\Omega$, 
remain unchanged, but the third parameter, the unperturbed stellar 
density~$\rho_0$ measured at the top~$Z_{\rm top}$ of the tower, changes. 
In fact, it drops rather rapidly for a typical collapsar progenitor.
Correspondingly, the radius of the tower increases as the tower grows.

Also notice that, by assumption, the pressure spreads rapidly along 
the cocoon. Therefore, the characteristic magnetic field and hence 
the radius of the tower are more or less constant along the tower 
at any given time. This justifies the assumption that the tower is 
a straight cylinder. In a more general situation, when the pressure 
equilibration in the cocoon lags somewhat, the radius of the tower 
will be a function of the vertical coordinate~$z$.
However, as is seen from equation~(\ref{eq-R0}), $R_0$ scales only 
weakly with~$\rho_0$ (as $\rho_0^{-1/6}$). This fact enables us to 
regard the constant density case as a good approximation.

%********************************************************************

\subsection{Let's Do the Numbers}
\label{subsec-numbers}

Now we will use the above relationships to make some quantitative
estimates based on the scaling relationships obtained in the previous
section; thus, unavoidably, these estimates can only provide
order-of-magnitude estimates.

We assume that the core of the star has collapsed into a black hole 
of fiducial mass $M=3\, M_{\odot}$ with a corresponding gravitational 
radius $R_g \equiv GM/c^2 \simeq 5\, {\rm km}$, and that some of the 
continuously infalling material has formed an accretion disk around 
the black hole. The pre-existing stellar magnetic field has been greatly 
amplified by the collapse because of flux-freezing and by the turbulent 
dynamo in the disk. Akiyama \& Wheeler (2003) have argued that the 
field may reach the level set by the equipartition with the MRI-driven
turbulence, as strong as $10^{16-17}\, \gauss$. This field is mostly
toroidal, however. The large-scale poloidal magnetic field that we need 
in our model may require a large-scale helical dynamo (Blackman~et~al. 2006) 
for its production and will probably be somewhat smaller than the toroidal 
field. Thus, we believe it is not unreasonable for the poloidal field at 
the disk surface to be a more modest $B_d\sim 10^{15}\, \gauss$ 
(see also Proga~et~al. 2003).

Now, if we take fiducial disk radius of $R_d \simeq R_{\rm ISCO}(a=0)= 
6 R_g \simeq 3\cdot 10^6 \, {\rm cm}$, where $a$ is the normalized Kerr
parameter of the black hole, we may have an initial poloidal 
dipole flux per unit toroidal angle of the order of $\Psi_0= R_d^2 B_0 
\simeq 10^{28} B_{15} R_{d,6.5}^2$ in cgs units.
(Note that if the black hole is spinning rapidly, then the inner disk radius 
is actually much closer, $R_{\rm ISCO}\simeq R_g \simeq 5\,{\rm km}$.)
This poloidal flux is being continuously twisted by the differential
Keplerian rotation of the disk, with characteristic angular velocity 
$\Delta\Omega = \Omega_K(R_d) \simeq 4\cdot 10^{3} {\rm sec^{-1}} 
(M/3M_{\rm Sun})^{1/2}\, R_{d,6.5}^{-3/2}$.

The typical background density of the stellar material into which the tower
propagates is $\rho_0=10^6\, {\rm g/cm^3}$. Hence, according to equation~(\ref
{eq-R0-Rd}), the tower radius can be estimated as
\beq
R_0 \sim R_d\, \biggl({{\tilde{V}_{A,d}}\over{V_d}}\biggr)^{1\over 3} =
3\, R_d\, \biggl({{B_{d,15}}
\over{R_{d,6.5}\Delta\Omega_{3.5}\sqrt{\rho_{0,6}}}}\biggr)^{1/3}\, ,
\label{eq-scaling-R_0}
\eeq
resulting in $R_0\simeq 10^7\, {\rm cm}$ for our fiducial parameter values.

Using this estimate for the outer radius of the tower, 
we can get the following expressions for all the other 
parameters:
\begin{eqnarray}
B_0 &\equiv& {\Psi_0\over{R_0^2}} = 
B_{\rm d} \biggl({R_{\rm d}\over{R_0}}\biggr)^2     \nonumber   \\
&\simeq&  0.1\, B_{\rm d}\, 
\biggl({{R_{d,6.5}\Delta\Omega_{3.5}}\over{B_{\rm d,15}}}\biggl)^{2/3}\, 
\rho_{0,6}^{1/3} \simeq 10^{14}\, \gauss           
\label{eq-scaling-B_0}                \, ; \\ 
V_{A,0} &\equiv& {B_0\over{\sqrt{4\pi\rho_0}}} =
3\cdot 10^{10}\, {\rm cm/sec}\ B_{d,15}^{1/3}\, R_{d,6.5}^{2/3}\,
\Delta\Omega_{3.5}^{2/3}\, \rho_{0,6}^{-1/6}    
\label{eq-scaling-V_A}            \, .
\end{eqnarray}

Notice that our crude estimate results in $V_{\rm top}\sim V_{A,0}$ 
being comparable to the speed of light~$c$; moreover, as we shall
show in~\S~\ref{sec-examples}, a growth velocity of $V_{\rm top}\simeq 
2.5 V_{A,0}$ is actually more typical. These facts strongly suggest
that a fully-relativistic treatment of the problem would be more 
appropriate (see \S~\ref{subsec-rel} for discussion). Such a treatment 
lies beyond the scope of the current paper, but we do intend to develop 
the relativistic magnetic tower theory in a subsequent paper.

Also, we can estimate the post-shock pressure in the hot cocoon 
above the tower as
\beq
P_{\rm top} \simeq {{B_0^2}\over{8\pi}} \simeq 
4\cdot 10^{26}{\rm erg\ cm^{-3}}\ B_{0,14}^2 \, .
\eeq
where $B_{0,14}\equiv B_0/(10^{14}\gauss)$.
At such very high energy densities the radiation pressure most likely
dominates over the gas pressure; we can therefore estimate the plasma
temperature in the post-shock region as
\beq
T_{\rm top} \simeq \biggl({3P_{\rm top}\over a}\biggr)^{1/4} \simeq
2\cdot 10^{10}\, {\rm K} \simeq 2\, {\rm MeV}\, ,
\eeq
where $a\simeq 7.6\cdot 10^{-15}\, {\rm erg\, cm^{-3}\, K^{-4}}$.
On the other hand, since we are dealing with a strong hydrodynamic
shock between the cocoon and the unperturbed stellar material, the 
baryon density in the cocoon is simply $4\rho_0$, and so the 
baryon rest-mass energy density is 
$4\rho_0 c^2 \simeq 4\cdot 10^{27}{\rm erg\ cm^{-3}}\,\rho_{0,6}$,
and hence still exceeds the radiation/pair energy density by an 
order of magnitude.

The total magnetic energy contained in the tower of height~$Z_{\rm top}$
can be estimated as 
\beq
E_{\rm mag}(t) \simeq 2\pi R_0^2\, Z_{\rm top}(t)\, {{B_0^2}\over{8\pi}}
\simeq 2\cdot 10^{50}\, {\rm erg}\ R_{0,7}^2\, Z_{\rm top,9}\, B_{0,14}^2 \, ,
\label{eq-Emag}
\eeq
which represents a noticeable fraction of a typical GRB energy 
(the factor~2 is added to take into account the fact that we 
actually have two towers, one above and one below the disk midplane).

%***************************************************************

\section{Mathematical Formalism}
\label{sec-equations}

The estimates presented in the previous section are very rough.
A more rigorous mathematical model is needed to demonstrate how
magnetic towers work and to obtain more accurate quantitative
estimates of the tower growth parameters. We shall proceed to
construct such a model. In particular, we shall develop a general 
analysis in this section and then use it in \S~\ref{sec-examples} 
to illustrate our model by specific examples.

Consider a magnetic tower making its way through a star. 
Since the system is axisymmetric, the magnetic field can 
be written in cylindrical polar coordinates~$(R,\phi,z)$ 
as
\beq
{\bf B}(R,z) = {\bf B}_{\rm pol} + B_\phi \hat{\phi} = 
{1\over R}\, [\nabla\Psi\times\hat{\phi}]+{I_{\rm pol}\over{R}}\,\hat{\phi}\, ,
\label{eq-B}
\eeq
where $\Psi(R,z)$ is the poloidal magnetic flux function,
$I_{\rm pol}(R,z)$ is $(2/c)$ times the poloidal electric 
current, and~$\hat{\phi}$ is the toroidal unit vector. 
If the magnetic field inside the tower is force-free, 
$I_{\rm pol}$ is constant along poloidal field lines, 
$I_{\rm pol}(R,z)=I_{\rm pol}(\Psi)$.

We shall focus on the middle part of Lynden-Bell's magnetic tower 
(his section~II), where the poloidal magnetic field is nearly straight 
and vertical. For simplicity we shall assume that the outer radius of 
the tower, $R_0$, is independent of height~$z$. This means that, at any 
given moment of time~$t$, all magnetic quantities depend only on one 
coordinate --- the cylindrical radius~$R$. Then, the force-free 
Grad--Shafranov equation --- the main {\it partial} differential 
equation (PDE) that governs the field structure  --- reduces to 
a 2nd-order {\it ordinary} differential equation (ODE) for the 
function~$\Psi(R)$:
\beq 
R\, \partial_R \biggl({1\over R}\, \Psi_R \biggr) = 
-\, I_{\rm pol} I_{\rm pol}'(\Psi) \, ,
\label{eq-GS-1}
\eeq
supplemented by two boundary conditions,
\beq
\Psi(R=0) = \Psi_0 = \Psi(R=R_0) \, .
\label{eq-bc-1}
\eeq

As a side remark, we could consider a more general situation where 
the outer radius slowly changes with height, $R_0=R_0(z)$. Then, 
instead of just one ODE, we would have a one-parameter infinite 
set of ODEs, one for each height~$z$. All these equations would 
have the same form as equation~(\ref{eq-GS-1}); the local height~$z$ 
would come in indirectly through the outer boundary condition $\Psi=\Psi_0$ 
at~$R=R_0(z)$. 

Equation (\ref{eq-GS-1}) involves a free function, 
the poloidal current~$I_{\rm pol}(\Psi)$. In reality,  
this function is not arbitrary but is determined by 
the differential rotation of the footpoints. 
By a simple geometrical consideration, the twist 
angle~$\Delta\Phi(\Psi)$ (i.e., $2\pi$ times the 
number of rotations) can be written as 
\beq
\Delta\Phi(\Psi) = 
I_{\rm pol}(\Psi)\, \int\limits_\Psi {{dl_{\rm pol}}\over{R^2 B_{\rm pol}}} 
= I_{\rm pol}(\Psi)\, \int\limits_\Psi {{dz}\over{R^2 B_z}} \, ,
\label{eq-twist-general}
\eeq
where the integration is performed along the field line and $l_{\rm pol}$ 
is the path length along the poloidal field.
Assuming that most of the twist occurs in the main part
of the magnetic tower, where the poloidal field is nearly 
vertical, we can write:
\beq
\Delta\Phi (\Psi) = I_{\rm pol}(\Psi)\, Z_{\rm top}(t)\, 
\biggl[{1\over{R_1^2(\Psi) |B_{z1}(\Psi)|}} + 
{1\over{R_2^2(\Psi) |B_{z2}(\Psi)|}} \biggr] \, ,
\label{eq-twist-tower}
\eeq
where $Z_{\rm top}(t)$ is the total height of the tower; 
its time-dependence reflects the fact that the tower grows 
with time. Here, $R_1(\Psi)$ and~$R_2(\Psi)$ are the cylindrical 
radii of the two segments of the flux surface~$\Psi$, and 
$B_{z1}(\Psi)$ and $B_{z2}(\Psi)$ are the values of the 
vertical magnetic field on these segments.

On the other hand, the twist angle is determined by the disk rotation law,
\beq
\Delta\Phi(\Psi) = \Delta\Omega(\Psi) t \, .
\label{eq-rotation}
\eeq
where~$\Delta\Omega(\Psi)$ is the difference between the angular 
velocities of the two disk footpoints of the field line~$\Psi$.
Thus, determining the twist angle really requires the knowledge
of both the radial disk rotation profile, $\Omega_d(r)$, and the 
radial distribution of the poloidal magnetic flux on the surface 
of the disk, $\Psi_d(r)$. Whereas there exists a very natural choice 
for the first of these two functions --- the Keplerian rotation law, 
$\Omega_d(r)=\Omega_K(r) \sim r^{-3/2}$, the second function, i.e, 
$\Psi_d(r)$, is not very well known. In reality it is determined by 
complicated MHD processes of turbulent dynamo and turbulent magnetic
flux transport, which are presently understood only poorly and 
definitely lie beyond the scope of the this paper. 
Therefore, instead of pretending that we know how~$\Psi_d(r)$ is really 
determined, we shall regard this function as essentially arbitrary. 
Correspondingly, instead of specifying it explicitly, we shall choose 
an indirect prescription dictated mostly by our mathematical convenience. 
In particular, we shall take~$\Psi_d(r)$ to be such as to result in a 
convenient and simple functional form for~$I_{\rm pol}(\Psi)$. 
Thus, we shall use the following scheme: first, we shall pick 
a function~$I_{\rm pol}(\Psi)$ that will make our calculations 
easier; then, we shall solve the Grad--Shafranov equation for 
this choice of~$I_{\rm pol}(\Psi)$; and finally, we shall use
equation~(\ref{eq-twist-tower}) to find~$\Delta\Omega(\Psi)$ 
and hence~$\Psi_{d}(r)$ a posteriori. This approach is similar 
to what is known as the ``generating-function method'' in the 
theory of force-free equilibria (e.g., Uzdensky~2002). 
Even though the main logic of such a scheme is physically backwards, 
it makes sense to use it as long as we don't get an unreasonable 
functional form for~$\Psi_{d}(r)$ in the end. 
At any rate, the alternative direct method, based on an 
explicit specification of~$\Psi_{d}(r)$ and on a subsequent 
use of equation~(\ref{eq-twist-tower}) to determine the 
corresponding~$I_{\rm pol}(\Psi)$, would be more difficult 
to implement and, at the same time, would also lack a good 
physical justification since the prescription for~$\Psi_{d}(r)$ 
would necessarily be quite arbitrary.

That said, there are some important features that this function is 
expected to have. For example, we expect the separatrix~$\Psi_s$ to 
correspond to the polarity inversion line on the disk surface, i.e., 
a line where $B_{z,d}$ goes through zero. In general, however, the 
first radial derivative of $B_z(r)$ is not zero, and so generically, 
we expect the disk poloidal flux function to behave as $\Psi_d\simeq \Psi_s 
+O[(r-r_s)^2]$ near this point. Since the first radial derivative of 
$\Omega_K(r)$ is finite at this point, we may expect 
\beq
\Delta\Omega (\Psi) \sim \sqrt{|\Psi-\Psi_s|}, \qquad 
{\rm as}\ \Psi\rightarrow\Psi_s \, .
\eeq

Next, we need to determine the radius $R_0$ of the tower.
This is done by equilibrating the inside magnetic pressure 
to the outside cocoon gas pressure at~$R_0$:
\beq
{{B_\phi^2(R_0)}\over{8\pi}}+{{B_z^2(R_0)}\over{8\pi}}= P_{\rm side}\, . 
\label{eq-side-equil}
\eeq

For simplicity we shall assume that the pressure is more or less
uniform throughout the side part of the cocoon and constitutes a
certain fixed fraction~$\eta$ of the gas pressure in the post-shock 
region directly above the tower: $P_{\rm side}=\eta P_{\rm top}$, 
$\eta\leq 1$. The latter, in turn, is determined from the condition 
of force balance across the contact discontinuity at the top of the 
magnetic tower. Following Lynden-Bell (1996), we ignore the 
complications resulting from the complex shape of this boundary 
and treat the tower top as a solid circular lid of radius~$R_0$. 
Then we  only need to satisfy the integral force balance, which 
can be written as
\beq
P_{\rm top}= {F_z\over{\pi R_0^2}} \, ,
\label{eq-P_top}
\eeq
where $F_z$ is the total magnetic force acting on the top of the tower. 
It can be computed by integrating the vertical flux of momentum associated 
with Maxwell's stress over the tower's cross-section:
\beq
F_z= 2\pi \int \limits_0^{R_0} 
\biggl[{{B_\phi^2(R)}\over{8\pi}}\,-\,{{B_z^2(R)}\over{8\pi}} \biggr] RdR \, .
\label{eq-F_z-1}
\eeq

Using equation~(\ref{eq-B}), we can rewrite this as
\beq
F_z= {1\over 4}\, \int \limits_0^{R_0} 
\biggl[I_{\rm pol}^2 - (\partial_R\Psi)^2 \biggr]\, {dR\over R} \, .
\label{eq-F_z-2}
\eeq

Finally, once the pressure at the top of the tower is known, 
we can use equation~(\ref{eq-shock-pressure}) to estimate the
speed at which the tower expands vertically,
\beq
V_{\rm top}=\dot{Z}_{\rm top} =
\sqrt{{3\over 4}{P_{\rm top}\over{\rho_0}}} \, .
\label{eq-V_top}
\eeq

%----------------------------------------------------------------

\subsection{Formulation of the Problem in Dimensionless Variables}
\label{subsec-dimensionless}

In the following, it will be convenient to rescale $R$ 
by the tower radius~$R_0$, poloidal flux $\Psi$ 
by the total flux in the tower~$\Psi_0$, and poloidal 
current $I_{\rm pol}$ by $\Psi_0/R_0$. Thus we 
define new dimensionless variables
\begin{eqnarray}
x  &\equiv& {R\over{R_0}} \, , \\ 
\psi &\equiv& {\Psi\over{\Psi_0}} \, , \\
I &\equiv& {{I_{\rm pol} R_0}\over{\Psi_0}} \, .
\end{eqnarray}

It will also be useful to introduce a characteristic magnetic
field strength
\beq
B_0 \equiv {\Psi_0\over{R_0^2}} \, .
\eeq

The Grad-Shafranov equation~(\ref{eq-GS-1}) can be written in 
these dimensionless variables as
\beq
x\partial_x \biggl({1\over x}\, \psi_x \biggr) = -\, II'(\psi) \, .
\label{eq-GS-2}
\eeq

\begin{figure}
%\plotone{tower.eps}
\plotone{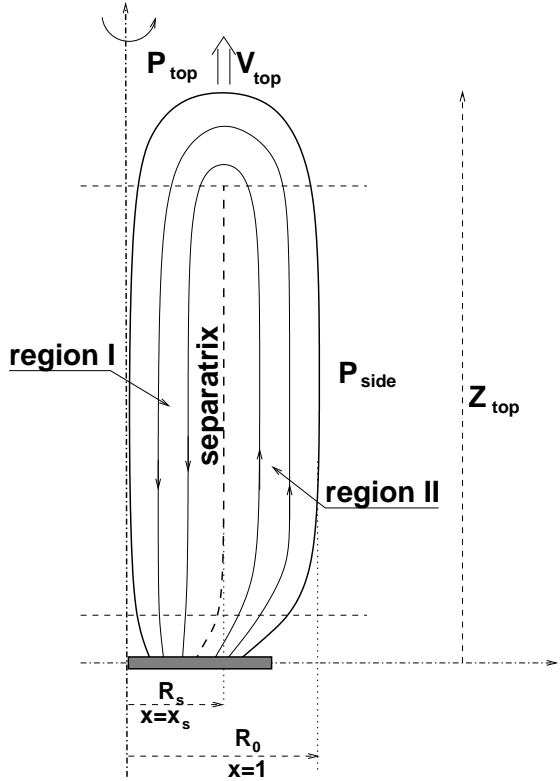}
\figcaption{A schematic drawing of a magnetic tower. The lines attached 
to the disk represent poloidal magnetic field lines (toroidal magnetic 
field is also present).
\label{fig-tower}}
\end{figure}

Geometrically, the magnetic tower consists of two regions
(see Fig.~\ref{fig-tower}): \\
1) inner region I of descending poloidal flux: $0 \leq x \leq x_s$; \\
and \\
2)  outer region II of ascending poloidal flux: $x_s \leq x \leq 1$.

The boundary between these two regions is the separatrix surface
$x=x_s$. By assumption, each field line leaving the disk and rising
up in region I will come back down to the disk through region~II;
therefore the total amount of poloidal flux in the two regions is
the same. We shall need to solve the Grad--Shafranov equation~(\ref{eq-GS-2}) 
separately in each of these two regions and then match the solutions at 
the separatrix.
Correspondingly, we set the boundary conditions as:
\begin{eqnarray}
\psi^I(x=0) &=& \psi^{II}(x=1) = 1  \label{eq-bc-Psi_tot} \, , \\
\psi^I(x=x_s) &=& \psi^{II}(x=x_s) = \psi_s      
\label{eq-bc-Psi_s}                          \, .
\end{eqnarray}
Without losing generality, from now on we shall set $\psi_s=0$.

The position of the separatrix $x_s$ is determined from the force-balance
condition across it:
\beq
(B_\phi^2 + B_z^2 )^I = (B_\phi^2 + B_z^2 )^{II}
\label{eq-sepx-equil-1}
\eeq

This condition can be greatly simplified, however. Indeed, first, 
because $B_\phi = I_{\rm pol}(\psi)/R$ and because the two sides
of the separatrix correspond to the same field line $\psi=\psi_s$,
and hence $I^{I}(\psi_s) = I^{II}(\psi_s)$, we 
have $(B_\phi^2)^I = (B_\phi^2)^{II}$. (Moreover, we shall 
actually assume that $I\rightarrow \sqrt{\psi} \rightarrow 0$ near
the separatrix, so $B_\phi$ actually goes to zero at the separatrix.)
Thus, our force-balance condition becomes:
\beq
(B_z^2 )^I = (B_z^2 )^{II} \Rightarrow 
\psi_x^I = -\psi_x^{II} \quad {\rm at } \ x=x_s \, .
\label{eq-sepx-equil-2}
\eeq

The vertical force $F_z$ is
\beq
F_z= {1\over 4}\, B_0^2 R_0^2\, 
\int\limits_0^1 \, [I^2(\psi)-\psi_x^2]\, {dx\over x} \, ,
\label{eq-F_z-dimensionless}
\eeq
and hence the pressure at the top of the tower is
\beq
P_{\rm top} =  {1\over{4\pi}}\, B_0^2 \, 
\int\limits_0^1 \, [I^2(\psi)-\psi_x^2]\, {dx\over x} \, ,
\label{eq-P_top-dimensionless}
\eeq

The condition~(\ref{eq-side-equil}) of pressure balance across 
the side wall of the tower, $x=1$, then becomes
\begin{eqnarray}
I^2(\psi=1)+\psi_x^2(x=1) &=& 8\pi\, {{P_{\rm side}}\over{B_0^2}} =
8\pi\eta \, {{P_{\rm top}}\over{B_0^2}} \nonumber \\
&=&2\eta\, \int\limits_0^1 \, [I^2(\psi)-\psi_x^2]\, {dx\over x} \, ,
\label{eq-side-equil-dimensionless}
\end{eqnarray}

Finally, we can rewrite the expression~(\ref{eq-twist-general})
for the field-line twist angle~$\Delta\Phi(\Psi)$ in our dimensionless
variables as
\beq
\Delta\Phi(\psi)=I(\psi)\, {{V_{\rm top}}\over{R_0}}\, t\, 
\biggl[{1\over{x_1|\psi_x(x_1)|}}+{1\over{x_2|\psi_x(x_2)|}}\biggr]\, ,
\label{eq-twist-dimensionless}
\eeq
where $x_1=x_1(\psi)$ and $x_2=x_2(\psi)$ are the radial positions
of the inner and outer segments of field line~$\psi$, respectively.

%***************************************************************

\section{Examples}
\label{sec-examples}

When dealing with any complicated physical system,
it is often useful to have at hand a few analytical examples that could
be used to explicitly illustrate the general characteristic behavior of
the system. In our particular problem, the simplest way to obtain explicit
analytical solutions is to consider the cases in which the Grad--Shafranov 
equation~(\ref{eq-GS-2}) becomes linear. The most general functional form 
of~$I(\psi)$ that leads to a linear Grad--Shafranov equation is 
\beq
I(\psi) = \sqrt{a\psi^2 + b\psi +c} \, .
\eeq

For simplicity, we shall consider the situation where the poloidal
current vanishes along the separatrix, $I(\psi=0)=0$, which makes
$c=0$. Other choices for $c$ are possible and may result in
interesting solutions. For now we consider $c=0$ and thus have 
\beq
I(\psi) = \sqrt{\psi} \, \sqrt{a\psi + b} \, .
\label{eq-I-psi-general}
\eeq

There are several interesting special cases that we will consider 
in this section.

As we shall see, these examples should be seen as mathematical
idealizations. In particular, the specific functional form of~$I(\psi)$, 
given by equation~(\ref{eq-I-psi-general}), may lead to solutions that 
are not regular at the axis. A physically-motivated regularity requirement 
at $R=0$ means that $I(\psi)$ should go to zero at the axis fast enough 
(e.g., like $1-\psi$).

Mathematically, solutions obtained in this section can be seen as 
describing the magnetic field behavior well outside of a narrow 
axial boundary layer, the inner ``core jet'', inside of which 
the poloidal current distribution~$I(\psi)$ deviates from equation~(\ref
{eq-I-psi-general}). The resulting singularity in our Case~2 in particular 
is similar in nature to the divergent behavior of the azimuthal magnetic 
field of a line current carried by a thin wire (inside of which the magnetic 
field is of course regular).

%******************************************************************

\subsection{Case 1: a Configuration without a Central Line Current}
\label{subsec-case-1}

First, we shall consider the case in which $I(\psi)$ behaves 
as~$\sqrt{\psi}$ near the separatrix~$\psi=0$ and at the 
same time goes to zero at the rotation axis: $I(1)=0$. 
Thus we choose $b=-a\equiv\mu^2>0$ in equation~(\ref{eq-I-psi-general}) 
and hence
\beq
I(\psi) = \mu \sqrt{\psi} \sqrt{1-\psi} \, ,
\eeq
which corresponds to $-II'(\psi)=\mu^2\, (\psi-1/2)$.

By substitutions $\psi=xu(x)+1/2$, this equation is reduced to 
the Modified Bessel's equation $x^2 u'' +x u' -u(\mu^2 x^2 +1)=0$,
and so the general solutions in the two regions can be written in 
terms of modified Bessel's functions as:
\begin{eqnarray}
\psi_I(x) &=& {1\over 2} + a_1 x I_1(\mu x)+ b_1 x K_1(\mu x) \, , \\
\psi_{II}(x) &=& {1\over 2} + a_2 x I_1(\mu x)+ b_2 x K_1(\mu x) 
\label{eq-solution2-general}     \, .
\end{eqnarray}

Here, $a_1$, $b_1$, $a_2$, and~$b_2$ are the arbitrary integration 
constants. They are determined by the boundary conditions~(\ref
{eq-bc-Psi_tot})--(\ref{eq-bc-Psi_s}) that can be cast as a linear 
system of four equations. Solving this system we get the explicit 
expressions for these coefficients in terms of the parameter~$\mu$ 
and the position of the separatrix~$x_s$:
\begin{eqnarray}
a_1 &=& -\, \mu\ {{1+y K_1(y)}\over{2yI_1(y)}} \, , \\
b_1 &=& {\mu\over 2} \, , \\
a_2 &=& -\, {1\over{2y}}\ 
{{yK_1(y) + \mu K_1(\mu)}\over{I_1(y)K_1(\mu)-K_1(y)I_1(\mu)}}\, ,\\
b_2 &=& {1\over{2y}}\ 
{{yI_1(y) + \mu I_1(\mu)}\over{I_1(y)K_1(\mu)-K_1(y)I_1(\mu)}} 
\label{eq-bc-case-1}             \, ,
\end{eqnarray}
where we defined  $y\equiv\mu x_s$ to simplify notation.

Next, the separatrix position~$x_s$ as a function of~$\mu$ 
is determined by the separatrix force-balance condition~(\ref
{eq-sepx-equil-2}) that can be written as
\beq
(a_1+a_2) I'_1(y) + (b_1+b_2) K'_1(y) = {\mu\over{y^2}} \, .
\label{eq-sepx-equil-case1}
\eeq
Thus, we get a non-trivial transcendental algebraic equation
that determines~$y$ as a function of~$\mu$. Once the solution of
this equation is found, we can use equations~(\ref{eq-bc-case-1}) 
to calculate the entire magnetic field structure. We solved 
equation~(\ref{eq-sepx-equil-case1}) numerically using {\it Mathematica}.
We found that the resulting dependence $y(\mu)$ is very close to linear 
$y=\mu/\sqrt{2}$ for small and finite values of~$\mu$. For $\mu\gg 1$ 
the slope of the linear dependence slightly changes and the asymptotic 
behavior becomes $y(\mu\rightarrow\infty)\simeq\mu-7$.
Correspondingly, the function $x_s(\mu)$ varies from $1/\sqrt{2}$ 
for $\mu\rightarrow 0$ to~1 for $\mu\rightarrow\infty$.
[The small-$\mu$ behavior is very easy to understand:
it corresponds to the situation in which the toroidal 
field is negligibly small, and so the horizontal force 
balance dictates that the magnitude of the vertical field
be uniform inside the tower. Since the total upward vertical 
flux  must be equal to the total downward flux, this means
that the area of region~I must be equal to that of region~II;
hence the separatrix between them has to be at~$x_s=2^{-1/2}$.]

Finally, we need to fix the value of~$\mu$. We do this by using 
the condition~(\ref{eq-side-equil}) of pressure balance between 
the magnetic field in the tower and gas pressure in the cocoon 
that surrounds and confines the tower. As discussed in \S~\ref
{sec-equations}, this condition can be viewed as the condition 
that determines the tower's radius once the poloidal current 
$I_{\rm pol}$ is specified. Here we, in effect, turn this around 
and determine~$\mu$ (which sets the overall scale for~$I_{\rm pol}$) 
in terms of the tower's radius. 

The solution of course depends on the assumed value of~$\eta$.
For a given~$\eta$, one can obtain a series of solutions for 
various values of~$\mu$ and then find the value $\mu_0(\eta)$ 
for which condition~(\ref{eq-side-equil}) is satisfied. We have 
done this using {\it Mathematica} and found that the allowable range
of~$\eta$ is limited. Namely, solutions exist only for~$\eta>\eta_c
\approx0.08$, with~$\mu_0(\eta)$ diverging as $(\eta-\eta_c)^{-1}$
in the limit~$\eta\rightarrow\eta_c$. Both~$P_{\rm top}$ and~$V_{\rm top}$ 
also diverge and $x_s$ approaches unity as~$\eta\rightarrow\eta_c$. 

\begin{figure}
\epsscale{1.05}
%\plotone{case1.eps}
\plotone{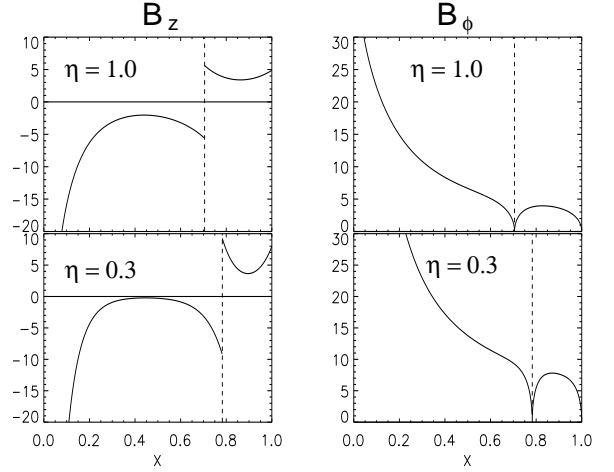}
\figcaption{Vertical (left) and toroidal (right) magnetic field components 
for two solutions in Case~1, computed for~$\eta=1.0$ (top panels) 
and $\eta=0.3$ (bottom panels). 
The vertical dashed lines on each plot shows the position~$x_s$ of 
the separatrix, across which the vertical field component ($B_z$) 
reverses sharply [the toroidal ($B_\phi$) component is symmetric 
with respect to~$x_s$].
\label{fig-case1}}
\end{figure}

We then investigated a few values of~$\eta$ in more detail. 
For $\eta=1$ we found $\mu_0(\eta=1)\simeq 6.70$ and 
$x_s(\eta=1)\simeq0.705$ (just a little less than~$1/\sqrt{2}$). 
The corresponding magnetic field components as functions of~$x$ 
are plotted in Figure~\ref{fig-case1} (the top two panels). 
We see that both the vertical and toroidal field components 
increase towards the axis, with $B_\phi\sim x^{-1/2}$.
At the separatrix (marked on the plots by the vertical dashed line),
the vertical field, as expected, reverses, whereas the toroidal field 
goes to zero on both sides of the separatrix as $|x-x_s|^{1/2}$.
The post-shock pressure above the tower is found to be $P_{\rm top}(\eta=1)
\simeq 0.945\, B_0^2$ and the vertical expansion speed of the tower, 
given by equation~(\ref{eq-V_top}), is $V_{\rm top}(\eta=1)\simeq 3\,V_{A,0}$.

It is interesting to note that, as can be seen in the top left panel 
of Figure~\ref{fig-case1}, the vertical magnetic flux tends to be 
pushed out (by the toroidal field pressure) from the intermediate-radii 
parts of each of the two tower regions ($x\sim 0.4-0.5$ in region~I and
$x\sim 0.8$ in region~II). As~$\eta$ decreases, this effect becomes even 
more pronounced, as can be seen from the bottom left panel of Figure~\ref
{fig-case1} which corresponds to~$\eta=0.3$. Also, in agreement with an
earlier discussion, the pressure at the top of the tower, $P_{\rm top}$, 
expressed in units of~$B_0^2$, and the tower growth velocity, $V_{\rm top}$, 
expressed in units of the Alfv\'en speed~$V_{A,0}$, both grow rapidly 
as~$\eta$ decreases. For example, in the case $\eta=0.3$, 
presented in the bottom two panels of Figure~\ref{fig-case1}, we 
find $P_{\rm top}(\eta=0.3)\simeq 8.4\,B_0^2$ and $V_{\rm top}(\eta=0.3)
\simeq 8.9\,V_{A,0}$.

%-------------------------------------------------------------------

\subsection{Case 2: a Configuration with a Central Line Current}
\label{subsec-case-2}

Let us now consider the case~$a=0$. Then, 
\beq
I(\psi)=\sqrt{b\psi} \equiv \lambda\sqrt{2\psi} \, .
\label{eq-I-psi-1}
\eeq

Notice that the poloidal current in this model does not vanish at 
the tower's rotation axis~$\psi=1$, implying the existence of a 
non-zero axial line current. Such a current may represent, for 
example, a very narrow core jet surrounded by the magnetic tower,
perhaps produced by the Blandford--Znajek (1977) mechanism acting 
along the field lines that thread the black hole. This line current 
may have important implications for the tower's growth speed and 
collimation, as we will discuss below.

The Grad--Shafranov equation~(\ref{eq-GS-2}) becomes:
\beq
x\, \partial_x \biggl({\psi_x\over x}\biggr)= -\lambda^2 ={\rm const}\, ,
\label{eq-GS-case2}
\eeq
and can be easily integrated:
\beq
\psi(x) = -\lambda^2\,{x^2\over 2}\,\log{x}+ C_1\, {x^2\over 2}+C_2\, .
\label{eq-solution1-general}
\eeq
This expression represents the general solution of the 2nd-order 
differential equation~(\ref{eq-GS-case2}). It involves two arbitrary 
integration constants, $C_1$ and $C_2$, that are 
determined by the boundary conditions~(\ref{eq-bc-Psi_tot})--(\ref
{eq-bc-Psi_s}). The resulting final expressions for $\psi(x)$ in 
the two regions are:
\begin{eqnarray}
\psi^I(x) &=& -\, {\lambda^2\over 2}\, x^2\, \log\biggl({x\over x_s}\biggr)+
1-{x^2\over{x_s^2}}
\label{eq-solution1-regI}    \, ,  \\
\psi^{II}(x) &=& -\, {\lambda^2\over 2} 
\biggl( x^2 \log x - {{1-x^2}\over{1-x_s^2}}\, x_s^2\, \log x_s \biggr) +
{{x^2-x_s^2}\over{1-x_s^2}}
\label{eq-solution1-regII}  \, .
\end{eqnarray}

\begin{figure}
%\plotone{lambda-xs.eps}
\plotone{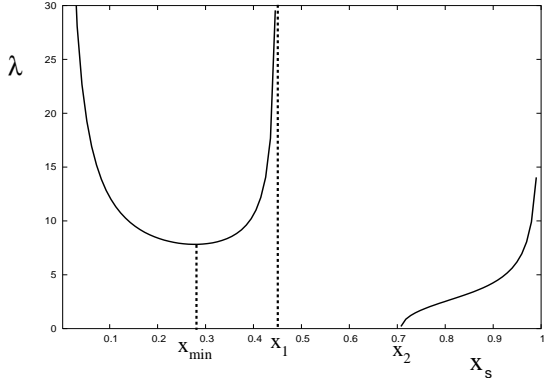}
\figcaption{The function $\lambda(x_s)$ corresponding to 
equation~(\ref{eq-lambda-xs}) for our Case~2.
\label{fig-lambda-xs}}
\end{figure}

Finally, the position $x_s$ of the separatrix is obtained by 
substituting these solutions into the separatrix force-balance 
condition~(\ref{eq-sepx-equil-2}), yielding a transcendental 
algebraic equation for~$x_s(\lambda)$. 
This equation can actually be resolved with respect to~$\lambda$ 
as a function of~$x_s$, resulting in an explicit expression:
\beq
\lambda(x_s) = {2\over{x_s}}\, 
\sqrt{{x_s^2-1/2}\over{1-x_s^2+\log x_s}} \, .
\label{eq-lambda-xs}
\eeq
This function is plotted in Figure~\ref{fig-lambda-xs}. 
One can see that there are two separate allowed ranges of~$x_s$: 
$0<x_s<x_1$, and $x_2 \leq x_s <1$, where $x_1\approx 0.45076$, 
and $x_2\equiv 1/\sqrt{2}\approx 0.707$. At $x_{\rm min}\approx 0.278$ 
the function has a minimum $\lambda(x_{\rm min})\approx~7.824$. 
For $x_s>x_2$, the solution~$\lambda(x_s)$ increases monotonically 
starting from $\lambda=0$ at $x_s=x_2$ and diverges as 
$\lambda\sim [2/(1-x_s)]^{1/2}$ as $x_s\rightarrow 1$.

%-------------------------------
%
%COMMENTED OUT: \\
%The asymptotic behavior of $\lambda(x_s)$ near the edges of 
%the allowed regions is described by
%\begin{eqnarray}
%\lambda(x_s\rightarrow 0)&\simeq& {\sqrt{2}\over{x_s\sqrt{|\log{x_s}|}}}\,;\\
%\lambda(x_s\rightarrow x_1) &\simeq& \sqrt{2\over{x_1(x_1-x_s)}} \simeq
%2.1\, (x_1-x_s)^{-1/2} \, ; \\
%\lambda(x_s\rightarrow x_2) &\simeq& 4\sqrt{\sqrt{2}\over{1-\log{2}}}\, 
%\sqrt{x_s-1/\sqrt{2}} \simeq 8.587\, (x_s-1/\sqrt{2})^{1/2} \, ; \\
%\lambda(x_s\rightarrow 1) &\simeq& \sqrt{2\over{1-x_s}} \, .
%\end{eqnarray}
%
%-----------------------------------

The next step in our program is to use the obtained solution 
to determine the vertical expansion velocity of the tower, 
$V_{\rm top}$. To do this, we first need to compute the 
integrated magnetic stress on the tower's top, given by 
equation~(\ref{eq-F_z-dimensionless}).
Notice, however, that in this model, as in any model with 
a non-zero line current along the axis, 
the toroidal magnetic field diverges as $1/x$ as $x\rightarrow 0$,
and hence its pressure diverges as $1/x^2$. Then, the 
vertical force due to the toroidal field pressure integrated 
over the top lid diverges logarithmically at small radii.
To manage this singularity, we will modify the problem somewhat by 
introducing a small inner cut-off radius $R_{\rm in}=x_{\rm in}R_0$ 
of the tower, $x_{\rm in}\ll 1$. This inner radius corresponds to 
the radius of the central jet that is embedded in the tower and 
that is actually responsible for carrying the axial line current. 
We use $x_{\rm in}$ as a small parameter to facilitate our
analysis.

The contribution of the toroidal field pressure to the overall integrated 
vertical force, 
\beq
F_z^{(B_\phi)} \equiv 2\pi R_0^2 \, \int\limits_{x_{\rm in}}^1 \, 
{{B_\phi^2}\over{8\pi}}\, xdx = {{\Psi_0^2}\over{4R_0^2}}\, 
\int\limits_{x_{\rm in}}^1\, I^2(\psi)\,{dx\over x} \, ,
\eeq
is dominated by the contribution near the cut-off radius; 
using the function~$I(\psi)$ given by equation~(\ref{eq-I-psi-1}), 
this contribution can be estimated as 
\beq
F_z^{(B_\phi)} = {\lambda^2\over 2}\, {{\Psi_0^2}\over{R_0^2}}\, 
\int\limits_{x_{\rm in}}^1 \, \psi(x)\, d\log{x} \simeq
{\lambda^2\over 2}\, \gamma B_0^2\, R_0^2 \, .
\label{eq-case2-F_z-Bphi}
\eeq
where we introduced a large parameter
\beq
\gamma \equiv |\log{x_{\rm in}}| \gg 1\, .
\label{eq-def-gamma}
\eeq

\begin{figure}
%\plotone{xs-eta-case2.eps}
\plotone{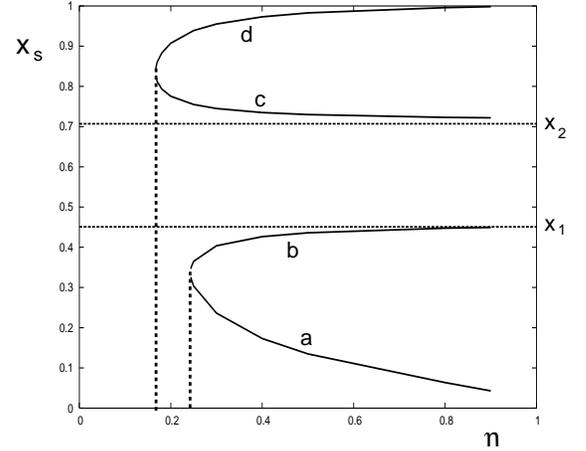}
\figcaption{The position $x_s$ of the separatrix as a function of~$\eta$ 
for the four solution branches corresponding to~$x_{\rm in}=10^{-4}$ in 
our Case~2.
\label{fig-xs-eta-case2}}
\end{figure}

Now, for what values of the parameters~$\eta$ and~$\gamma$ do solutions 
exist? To answer this question, we have used {\it Mathematica} to map out 
the two-dimensional ($\eta$,$\gamma$) parameter space. We have restricted 
our investigation to $\gamma\gg 1$ and~$\eta\leq 1$. We have found that 
the system exhibits a very rich mathematical behavior characterized by 
the multiplicity of solutions.
The overall picture is illustrated in Figure~\ref{fig-xs-eta-case2}, where 
we plot the separatrix position~$x_s$ as a function of~$\eta$ for a fixed 
value of~$\gamma=9.21$ (corresponding to $x_{\rm in}=10^{-4}$). We found
that essentially there are two pairs of solutions, one pair corresponding 
to~$x_s<x_2=2^{-1/2}$ and the other corresponding to~$x_s>x_2$. 
When $\eta=1$ there are only two solutions, one in each of these two regions. 
However, for $\eta<1$ two more solutions appear, so there are four in total, 
with very different properties.
As one decreases~$\eta$ at a fixed~$\gamma$, the difference between 
the two solutions in each pair gradually diminishes, until they finally 
merge into one solution. As can be seen from Figure~\ref{fig-xs-eta-case2}, 
this happens at two critical values of~$\eta$, which we 
call~$\eta_{\rm min,1}$ for the two solutions below~$x_2$ 
and~$\eta_{\rm min,2}$ for the two solutions above~$x_2$.
We found that both $\eta_{\rm min,1}$ and $\eta_{\rm min,2}$ scale 
inversely with~$\gamma$: $\eta_{\rm min,1}(\gamma)\approx 2.25\gamma^{-1}$,
and $\eta_{\rm min,2}(\gamma)\approx 1.5\gamma^{-1}$. 
There are no solutions for $\eta<\eta_{\rm min,1,2}(\gamma)$.

\begin{figure}
%\plotone{case2.eps}
\plotone{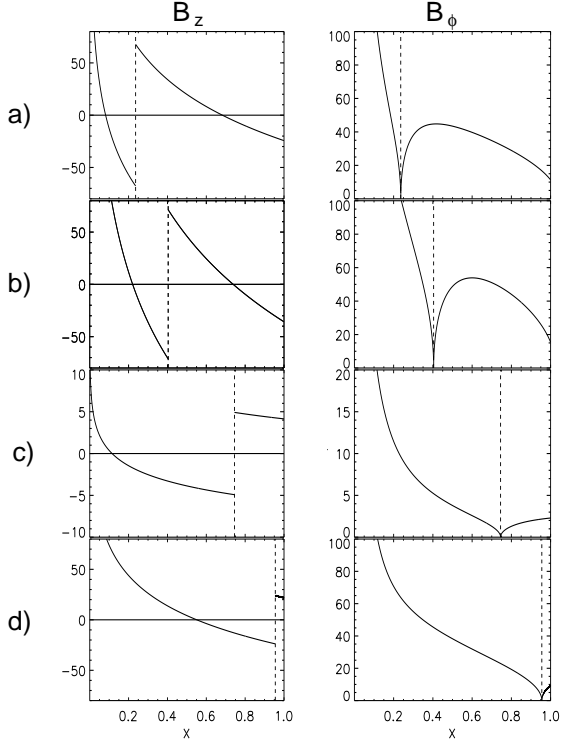}
\figcaption{Vertical (left) and toroidal (right) magnetic field components 
of the four solutions in Case~2, computed for $x_{\rm in}=10^{-4}$ 
($\gamma\simeq 9.2$) and~$\eta=0.3$. 
The vertical dashed lines on each plot shows the position~$x_s$ of 
the separatrix, across which the vertical field component ($B_z$) 
reverses sharply [the toroidal ($B_\phi$) component is symmetric 
with respect to~$x_s$].
\label{fig-case2}}
\end{figure}

As an illustration, we present the four solutions corresponding to 
$x_{\rm in}=10^{-4}$ ($\gamma=9.21$) and~$\eta=0.3$ obtained using 
{\it Mathematica}. In Figure~\ref{fig-case2} we plot the vertical 
and toroidal components of the magnetic field for these solutions.
One can see that the toroidal field goes to zero as $|x-x_s|^{1/2}$
on both sides of the separatrix without changing sign and grows
as~$x^{-1}$ near the axis, which is explained by the assumed shape
of the function~$I(\psi)$.
One can also notice that the vertical field in these solutions 
not only reverses sharply across the separatrix, but also changes 
sign by going smoothly through zero at one or two other points. 
This means that the magnetic field described by these solutions 
actually has a more complex topology than that of the simple tower 
of Figure~\ref{fig-tower}. Thus, solutions~2c and~2d correspond to 
two towers, one inside the other, and solutions~2a and~2b 
corresponds to three nested towers.

The reason for the existence of two solutions with $x_s>x_2$ 
for~$\eta>\eta_{\rm min,2}$ can be traced to the following argument. 
The vertical force~(\ref{eq-case2-F_z-Bphi}) leads to a cocoon side 
pressure that scales as $\lambda^2\gamma\eta B_0^2$. This pressure has 
to be balanced at the outer edge of the tower by the internal magnetic 
pressure that generally is of the order of~$B_0^2/(1-x_s)^2$. It turns 
out that, provided that $\eta>\eta_{\rm min,2}$, there are two ways to 
make such a balance, two different regimes, distinguished by whether~$x_s$ 
is close to~1 or not. We shall now consider these two possibilities 
(which we call Case~2c and Case~2d) separately and give basic analytical 
derivations under the assumption that $\eta\gamma\gg 1$ (and hence 
$\eta\gg\eta_{\rm min,2}$).

%-----------------------------------------------------------

{\bf Case 2c:} Let us assume that $x_s-1$ is finite. 
Then, the force balance condition at the side wall of the tower
can be satisfied only if~$\lambda$ is small,  namely, $\lambda^2
\propto(\eta\gamma)^{-1}$. This suggests the following overall 
picture. Since $\lambda^2$ is the factor that parameterizes the 
general strength of the toroidal field pressure, we see that the 
toroidal field term~$-II'(\Psi)$ in the Grad--Shafranov equation 
is unimportant in most of the tower, with the exception of a small 
vicinity of the inner edge. The field is then predominantly vertical 
almost everywhere; the force-free balance then dictates that this 
field must be essentially uniform in each of the two regions. Since 
the total upward-directed magnetic flux in region~II must equal to 
the total downward-directed flux in region~I, the areas of the two 
regions must then be the same, hence $x_s^2\approx 1/2$. Then we can 
approximate~$x_s$ in equations~(\ref{eq-solution1-regI})--(\ref
{eq-solution1-regII}) by $x_2=2^{-1/2}$ and rewrite them as
\begin{eqnarray}
\psi^I(x) &\simeq & 1-2 x^2 
\label{eq-solution1c-regI}  \, , \\
\psi^{II}(x) &\simeq& 2x^2-1 
\label{eq-solution1c-regII} \, ,
\end{eqnarray}
corresponding to $B_z^I(x)\simeq-\, 4B_0$, and $B_z^{II}(x)\simeq 4B_0$.

Next, the overall integrated force~$F_z$ is the sum of the positive 
contribution~$F_z^{(B_\phi)}$ of the toroidal field pressure near 
the axis and the negative contribution from the poloidal field 
tension in the rest of the tower:
\beq
F_z=F_z^{(B_\phi)} - \, {1\over 4}\, B_0^2 R_0^2\, 
\int\limits_{x_{\rm in}}^1\, \psi_x^2\, {dx\over x} \simeq
\biggl({{\lambda^2\gamma}\over 2}-2 \biggr)\, B_0^2 R_0^2 \, .
\label{eq-case2c-Fz}
\eeq
Then, using the condition~(\ref{eq-side-equil-dimensionless}) of 
the pressure balance between the tower and the cocoon at the outer 
edge~$x=1$, we obtain a simple approximate expression for~$\lambda$ 
in terms of~$\gamma$:
\beq
\lambda \simeq 2\,\sqrt{{1+\eta}\over{\eta\gamma}} \ll 1 \, ,
\label{eq-case2c-lambda-gamma}
\eeq
which reduces to $\lambda\simeq(8/\gamma)^{1/2}$ for~$\eta=1$
and $\lambda\simeq 2(\eta\gamma)^{-1/2}$ for~$\eta\ll 1$.
By combining this expression with the asymptotic behavior 
\beq
\lambda(x_s\rightarrow x_2) \simeq 4\sqrt{\sqrt{2}\over{1-\log{2}}}\, 
\sqrt{x_s-1/\sqrt{2}} \simeq 8.587\, (x_s-1/\sqrt{2})^{1/2} \, , 
\label{eq-lambda-xs-asymptotic}
\eeq
of the function $\lambda(x_s)$ [obtained from equation~(\ref{eq-lambda-xs})
in the limit $x_s\rightarrow~x_2$], we can determine exactly how close~$x_s$ 
has to be to~$x_2$ in terms of~$\gamma$ and $\eta$:
\beq
x_s-x_2 = {{1+\eta}\over{\eta\gamma}}\, {{1-\log{2}}\over{4\sqrt{2}}} 
\simeq 0.054\, {{1+\eta}\over{\eta\gamma}}      \, .
\label{eq-case2c-xs-gamma}
\eeq
This becomes $x_s-x_2\simeq 0.11\gamma^{-1}$ for~$\eta=1$
and $x_s-x_2\simeq 0.054\, (\eta\gamma)^{-1}$ for~$\eta\ll 1$. 

Finally, by substituting equation~(\ref{eq-case2c-lambda-gamma}) 
into~(\ref{eq-case2c-Fz}) and using relationship~(\ref{eq-P_top}),
we get an estimate for the cocoon pressure at the top of the tower:
\beq
P_{\rm top}\simeq {2\over{\pi\eta}}\,B_0^2\, \simeq{0.637\over\eta}\,B_0^2\,,
\label{eq-case2c-P_top}
\eeq
which is of course consistent with the pressure balance across 
the side wall between the cocoon and the magnetic field~$4B_0$
inside the tower. Equation~(\ref{eq-V_top}) can then be used to 
estimate the tower's vertical expansion velocity:
\beq
V_{\rm top}\simeq \sqrt{6/\eta}\,V_{A,0} \simeq 2.45\,\eta^{-1/2} V_{A,0}\, .
\label{eq-case2c-V_top}
\eeq
This result, together with its counterpart from the previous 
example, indicates that $V_{\rm top}$ between~2.5 and~3 $V_{A,0}$ 
provides a more accurate estimate for the tower growth velocity 
than equation~(\ref{eq-V_top=V_A}).

With these results at hand, we can go back and compute 
the differential rotation profile~$\Delta\Omega(\psi)$ 
that corresponds to this case. 
As long as we are not too close to~$\psi=1$, we can 
use equations~(\ref{eq-solution1c-regI}) 
and~(\ref{eq-solution1c-regII}) to explicitly 
express~$x_1(\psi)$ and~$x_2(\psi)$ in terms of~$\psi$ 
and we can also approximate $\psi_x^I=-\,4x$, $\psi_x^{II}=4x$. 
By substituting these approximations, along with~(\ref{eq-I-psi-1}),
into equation~(\ref{eq-twist-dimensionless}), we get a very simple 
relationship
\begin{eqnarray}
\Delta\Omega(\psi) = {\Delta\Phi\over t} &=& \sqrt{2}\lambda\, 
{{V_{A,0}}\over{R_0}}\, {{\sqrt{\psi}}\over{1-\psi^2}} \nonumber \\
&\simeq& {4\over{\sqrt{|\log{x_{\rm in}}|}}}\, \sqrt{{1+\eta}\over{2\eta}} 
\, {{V_{A,0}}\over{R_0}}\, {{\sqrt{\psi}}\over{1-\psi^2}} \, .
\label{eq-case2c-DeltaOmega}
\end{eqnarray}
This expression breaks down near $\psi=1$ as of course it should
since the toroidal field's pressure becomes significant and modifies
the poloidal field structure near the inner edge. Also note that the 
factor $\sqrt{2}\lambda\simeq 4|\log{x_{\rm in}}|^{-1/2}$ that is 
in front of this expression, although formally vanishing in the limit 
$x_{\rm in}\ll 1$, in reality is never very small; for example, it is 
equal to~$\approx~1.1$ for $x_{\rm in}=1\times 10^{-6}$.

We have used {\it Mathematica} to obtain exact (i.e., without relying 
on the assumption $\gamma\gg 1$) solutions for this case for several 
values of~$\eta$ and~$\gamma$. In particular, we found that, in the 
limit of large~$\gamma$, the resulting $\gamma$-dependences of~$\lambda$, 
the separatrix position~$x_s$, and the pressure in the cocoon~$P_{\rm top}$ 
at a fixed~$\eta$ are in a good agreement with the above simple estimates.
In addition, Figure~\ref{fig-case2}c shows the radial profiles 
of the vertical and toroidal magnetic field components inside 
the tower for a representative case $x_{\rm in}=0.0001$ and~$\eta=0.3$. 
One can see that the vertical field is indeed close to a constant~$4B_0$ 
outside the separatrix and $-\,4B_0$ inside, as expected, although 
it deviates significantly from this constant close to the central axis.

%---------------------------------------------------------------------

{\bf Case 2d:} Another interesting case is that corresponding 
to~$x_s$ being close to~1 and hence a large~$\lambda$. Then, 
assuming that the contribution~(\ref{eq-case2-F_z-Bphi}) of 
the toroidal pressure to the integrated vertical force 
dominates over the poloidal field contribution (this assumption
is valid as long as~$\eta\ll 1$, as we shall show below), 
we can estimate the post-shock pressure at the tower top as
\beq
P_{\rm top} = {{F_z^{(B_\phi)}}\over{\pi R_{\rm out}^2}} = 
{\lambda^2\over{2\pi}}\, B_0^2\, \gamma \, .
\label{eq-case2d-P_top}
\eeq

Applying the condition of pressure balance across~$R=R_0$, we then get:
\beq
{{I_{\rm pol}^2(\psi=1)}\over{R_0^2}} +
{{\Psi_0^2}\over{R_0^4}}\, \psi_x^2(x=1) = 8\pi P_{\rm side} =
8\pi \eta P_{\rm top}  = 4\lambda^2 \gamma \eta \, B_0^2\, .
\label{eq-case2d-equilibrium-side-1}
\eeq
The contribution of the toroidal field to the pressure balance 
at the outer boundary is $B_\phi^2(x=1)=I_{\rm pol}^2(\psi=1)/R_0^2= 
2\lambda^2\, B_0^2$. Assuming that $\eta\gamma\gg 1$, we can therefore
neglect it in comparison with $8\pi P_{\rm side}$, and so the above 
pressure balance condition yields
\beq
\psi_x(x=1) \simeq 2\lambda\, \sqrt{\gamma\eta} \, .
\label{eq-case2d-equilibrium-side-2}
\eeq
Now we want to use this relationship in conjunction with our general 
solution derived above to determine the parameter~$\lambda$. Notice 
that the introduction of the inner radius does not alter the general 
functional form of~$\psi(x)$ given by equation~(\ref{eq-solution1-general}).
The inner boundary condition is changed into $\psi(x_{\rm in})=1$. 
However, we expect that, for~$x_{\rm in}\ll 1$, the relationships  
between the coefficients that describe the function~$\psi(x)$ are 
changed only slightly, even though they had been derived using the 
boundary condition $\psi(x=0)=1$. Then, as long as we are not too 
close to the inner boundary, we can still use expressions~(\ref
{eq-solution1-regI})--(\ref{eq-solution1-regII}) for our solution. 
In particular, we can write $\psi_x(x=1)$ as
\beq
\psi_x(x=1) = {2\over{1-x_s^2}} - \lambda^2\, 
{{x_s^2}\over{1-x_s^2}}\, \log{x_s} -{\lambda^2\over 2} \, .
\eeq
We can simplify this expression by noticing that the relationship~(\ref
{eq-lambda-xs}) between~$\lambda$ and~$x_s$ can be cast in the form
$\lambda^2 x_s^2\log{x_s}=2(2x_s^2-1)-\lambda^2 x_s^2 (1-x_s^2)$.
Then, the above expression for $\psi_x(x=1)$ greatly simplifies 
and, substituting it into the pressure balance condition~(\ref
{eq-case2d-equilibrium-side-2}), we obtain
\beq
\psi_x(x=1) = 4 + \lambda^2\, \biggl(x_s^2-{1\over 2}\biggr) = 
2\lambda\, \sqrt{\gamma\eta} \, .
\label{eq-case2d-equilibrium-side-3}
\eeq
From this, we can express $x_s$ in terms of $\lambda$ as
\beq
x_s^2= {1\over 2}+2\, {{\lambda\sqrt{\gamma\eta}-2}\over{\lambda^2}} \, .
\eeq
This equation, together with equation~(\ref{eq-lambda-xs}), 
completely determines both~$x_s$ and~$\lambda$ for a given 
value of the product~$\gamma\eta\gg 1$. We can go even further 
and get explicit formulae for~$x_s$ and~$\lambda$ by making use
of the fact that~$x_s\approx 1$ [and hence, according to eq.~(\ref
{eq-lambda-xs}), $\lambda\gg 1$] in the regime under consideration. 
Then, the first term in~(\ref{eq-case2d-equilibrium-side-3}) is small 
compared with the second term and we immediately obtain
\beq
\lambda\simeq 4\sqrt{\gamma\eta}\equiv 4\sqrt{\eta|\log{x_{\rm in}}|}\gg 1 \,,
\label{eq-case2d-lambda-gamma}
\eeq

\beq
|\log{x_s}| \simeq 1-x_s \simeq 
{2\over{\lambda^2}} \simeq {1\over{8\gamma\eta}} \ll 1 \, .
\label{eq-case2d-x_s-gamma}
\eeq

With these results, we can now go back to our solution given by 
equations~(\ref{eq-solution1-regI})--(\ref{eq-solution1-regII})
and derive the following asymptotic expressions for~$\psi(x)$ in 
the two regions 
\begin{eqnarray}
\psi^I(x) &\simeq&
1 - 2x^2 - 8\gamma\eta\, x^2\, \log{x} \, , \qquad x\gg x_{\rm in} 
\label{eq-soln1d-regI}       \, ; \\
\psi^{II}(x) &\simeq&
1 - 8\gamma\eta\, (1-x^2 + x^2\, \log{x}) \simeq 1 - 8\gamma\eta\, (1-x) 
\label{eq-soln1d-regII}       \, .
\end{eqnarray}

From our estimations of the vertical force 
and the post-shock pressure above the tower, 
we find
\beq
F_z \simeq
{{\gamma\lambda^2}\over 2}\, B_0^2\, R_0^2 \simeq 
8 \gamma^2\eta \, B_0^2\, R_0^2 \, ,
\eeq

\beq
P_{\rm top} \simeq {{8\gamma^2}\over{\pi}}\, \eta B_0^2 \, .
\eeq
Correspondingly, the vertical expansion speed is 
\beq
V_{\rm top} = \sqrt{{3\over 4}\, {{P_{\rm top}}\over{\rho_0}}} \simeq
\sqrt{{6\eta}\over\pi}\, \gamma\, {B_0\over{\sqrt{\rho_0}}} =
\sqrt{24\eta}\, \gamma\, V_{A,0} \gg V_{A,0} \, .
\eeq
This suggests that magnetic tower configurations with a non-zero 
line current along the axis may be especially effective in producing 
jets that are rapidly propagating through a confining external medium.

As a final check, we now go back to our estimate of the integrated vertical 
force and ask under what circumstances it is indeed dominated by the toroidal 
field contribution. 
Using our solution~(\ref{eq-soln1d-regI})--(\ref{eq-soln1d-regII}) and
assuming that~$\gamma\eta\gg 1$, the poloidal field contribution can be
estimated as $F_z^{(B_z)}\simeq -\,8\gamma^2\eta^2\,B_0^2 R_0^2$, whereas,
substituting~(\ref{eq-case2d-lambda-gamma}) into 
equation~(\ref{eq-case2-F_z-Bphi}), one estimates the toroidal field 
contribution to be $F_z^{(B_\phi)}\simeq 8\gamma^2\eta\,B_0^2 R_0^2$.
Thus, we see that $F_z^{(B_\phi)}$ indeed dominates as long as~$\eta\ll 1$.

%-----------------------------------------------------------------

%****************************************************************

\section{Discussion}
\label{sec-discussion}

%-----------------------------------------------------------------

\subsection{Transition to the Relativistic Expansion Regime 
and the Final Opening Angle of the Jet}
\label{subsec-rel}

%[In general, we have to distinguish between relativistic
%expansion velocity and the relativistic field-line rotation
%velocity ($R_0\gg R_{\rm LC})$ inside the tower. In our
%non-relativistic analysis these two velocities are comparable,
%but it is not going to be so in the relativistic case.
%In addition to the expansion velocity, there is the issue of 
%relativistic rotation of field lines inside the tower.
%This happens when the radius of the tower is larger than
%the light cylinder so that $\Omega R_0 \gg c$.]

As we have shown in \S~\ref{subsec-cocoon-shock}, the radius~$R_0$ 
of the tower, and hence the rotational velocity of the field lines 
and the tower's growth 
velocity $V_{\rm top}\sim \Delta\Omega R_0$ scale with the background density
as~$\rho_0^{-1/6}$. Therefore, as the tower expands into the outer, less dense
regions of the star, $V_{\rm top}$ inevitably reaches the speed of light at a 
certain point that corresponds to some critical density~$\rho_{0,\rm rel}$.
Our non-relativistic model becomes becomes invalid at this point and any 
further expansion of the tower requires a relativistic treatment.
For our fiducial values $B_d= 10^{15}\ {\rm G}$, $R_d=3\cdot10^6\ {\rm cm}$, 
and $\Delta\Omega= 3\cdot 10^3\ {\rm sec^{-1}}$, this transition to the 
relativistic regime takes place at a critical density $\rho_{0,\rm rel}\sim
10^6\ {\rm g/cm^3}$ [see eq.~(\ref{eq-scaling-V_A})]. For a typical massive 
stellar GRB progenitor this density corresponds to a distance $Z_{\rm rel}$ 
from the center that is on the order of a few times $10^8\ {\rm cm}$; 
according to equation~(\ref{eq-scaling-R_0}), the radius of the tower 
corresponding to this critical density is $R_{\rm rel}\sim 10^7\ {\rm cm}$. 

A relativistic generalization of the magnetic tower model is beyond 
the scope of this paper but is something we intend to develop in the 
near future. Before a complete solution is achieved, however, we cannot 
really say anything definite about further expansion of the tower. 
However, we can, perhaps, look at the results of fully-relativistic 
{\it hydrodynamic} simulations (Zhang~et~al. 2003) for some physical insight. 
In those simulations the relativistic jet remained collimated as it 
propagated through the star, across several orders of magnitude in 
stellar density. The authors of that paper have attributed the 
observed collimation to recollimation shocks in the cocoon and 
to relativistic beaming in the jet. But the physical processes 
in the cocoon should not change if we replace the inner relativistic 
hydrodynamic jet with a relativistic magnetic tower. Furthermore,
we expect some additional, magnetic collimation due to the toroidal-field 
hoop stress (the pinch effect).
The bottom line is that we expect the magnetic tower to remain well 
collimated even after it transitions into the relativistic regime. 
In particular, we suggest that the final opening angle of the 
magnetically-dominated outflow will be no larger than the inverse 
aspect ratio of the tower at the moment of relativistic transition:
\beq
\Delta\theta \lesssim {{R_{0,\rm rel}}\over{Z_{\rm rel}}} \simeq 0.1 \, .
\eeq
Whether this prediction is true will have to be determined by
a fully-relativistic analysis and by relativistic MHD simulations,
which we hope will be completed in the near future. In addition,
we acknowledge that some first steps toward developing an analytical
understanding of magnetically-dominated outflows in the ultra-relativistic
regime were taken recently by Lovelace \& Romanova (2003) and by 
Lyutikov~(2006).

%*******************************************************************

\subsection{Tower Structure}
\label{subsec-structure}

Figure~\ref{fig-case1} shows the run of vertical and azimuthal magnetic 
field components for Case~1 as a function of scaled radius~$x$.  
For the $\eta = 1.0$ case (bottom panel), the toroidal field
scales as $x^{-1/2}$ so that the enclosed toroidal magnetic energy
increases linearly with radius.  The toroidal magnetic energy is not
concentrated toward the axis, but is distributed throughout the tower.
For Case~2 with a central line current (Figure~\ref{fig-case2}),
however, the toroidal magnetic field increases toward the axis ($x=0$)
as~$x^{-1}$.  The magnetic energy density is thus strongly peaked
toward the axis for this case.  This demonstrates that the presence of
a central line current results in a more tightly beamed outflow.

\begin{figure}
%\plotone{twotower.eps} 
\plotone{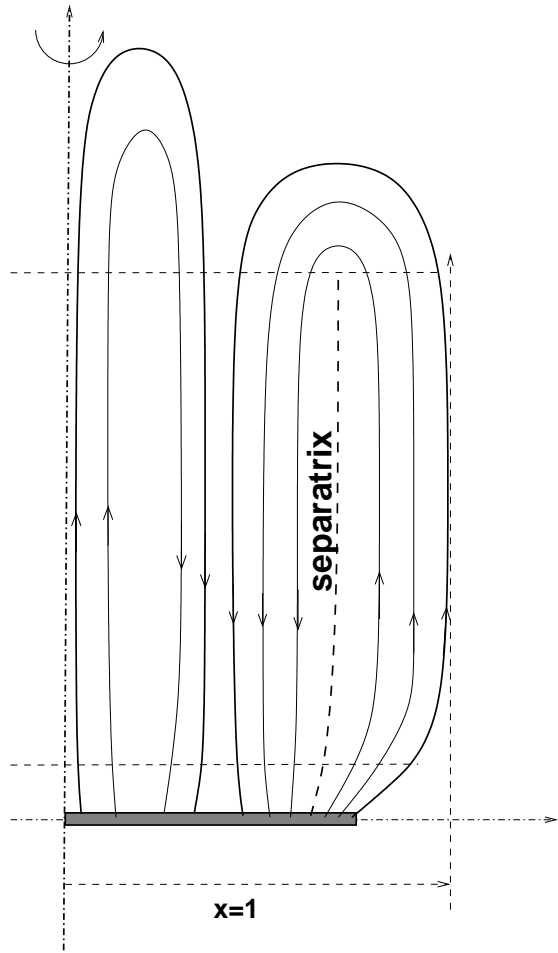} 
\figcaption{A schematic drawing of a twin magnetic
tower.  The outer tower which forms a hollow cylindrical shell which
surrounds the inner tower.
\label{fig-twintower}}
\end{figure}

For the field structures of Case 2 plotted in Figure~\ref{fig-case2}
we found solutions corresponding to multiple nested towers.  Case 2c,
for example, has B$_z$ passing through 0 at $x \sim 0.1$ and a
separatrix at $x \sim 0.75$ corresponding to two distinct towers, one
within the other.  Field lines in the inner tower are tied to the disk
deeper in the potential of the black hole or neutron star where the
characteristic velocity becomes comparable to the speed of light.
Thus, the inner tower may be a more energetic component of the outflow
and may reach the ultra-relativistic expansion regime earlier.  In
addition, the inner tower is shielded from the cocoon material by the
outer tower which envelopes it.  This is beneficial for achieving
relativistic flow in the inner tower since baryon contamination is
suppressed.  We note that, in principle, the tops of the towers can
propagate at different speeds, as indicated schematically by the tower
heights in Figure \ref{fig-twintower}.  However, we have not analyzed
this case in detail since all analysis we present here assumes a
single tower height.

The physical picture presented in this paper, with a smooth coherent
magnetic structure of the outflow, is an idealization, necessary for
obtaining a basic physical insight into the system's dynaimcs and to
get the main ideas across in the clearest way possible.
The actual magnetic field, especially if it is produced by a turbulent
dynamo in the disk, is likely to be different from such a simple system
of nested axisymmetric flux surfaces. Instead, it may consist of an
ensemble of loops of different sizes and orientations. It may thus
have a highly-intermittent substructure on smaller scales, both
temporal and spatial. In fact, our Case~2 solutions discussed above,
with two magnetic towers, one enveloped by the other, represent a
simple specific example of this general situation.

What is important though, is that each of these smaller magnetic
structures is subject to the same physical processes as a single
large tower: twisting due to differential rotation and a subsequent
inflation, controlled by the external pressure of the cocoon and of
other loops growing at the same time. As a result, a single tower may
be replaced by a train of spheromak-like plasmoids, pushing each other
out along the axis. The hoop stress still works inside each of them,
and so the overall dynamical effect may be the same as that of a single
tower, at least qualitatively. The same picture is expected to develop
if a single large-scale tower becomes unstable to reconnection and breaks
up into smaller plasmoids, as we discuss in Section~\ref{subsec-reconnection}.
In either case, the resulting multi-component structure of the outflow
may be responsible for the observed intermittancy in GRBs.

%*******************************************************************

\subsection{Baryon Contamination from the Disk}
\label{subsec-baryons}

The tower plasma may be polluted with baryons from the disk by two main
processes, magneto-centrifugal acceleration (Blandford \& Payne 1982) and
radiative neutrino-driven ablation (Qian \& Woosley 1996; Levinson~2006).  
In the case of
magneto-centrifugal acceleration, the field lines must be inclined outward by
at least 30 degrees from vertical for a disk wind to be launched.  It is of
interest to note that, for our configuration, field lines forming the inner
tower are plausibly constrained to leave the disk at progressively steeper
angles, moving inward toward the rotation axis.  This field configuration
would suppress the magneto-centrifugal launching mechanism for the inner
tower.  The outer tower may be more baryon loaded than the inner tower and
would constitute a less relativistic magnetically dominated flow at larger
angles from the pole.  Yet farther from the pole, the cocoon represents an
even higher baryon loaded component and is consequently expected to be slower
and contain a relatively small magnetization (whatever magnetic field is
mixed into the cocoon by plasma instabilities). Future relativistic
magnetohydrodynamical simulations will be able to address these issues and in
particular determine the angle of field lines emerging from the inner disk
enabling better estimates of baryon contamination.  In addition, we note that
in the case of black hole--disk magnetic coupling, one should expect no
baryon loading at all along the part of a field line that connects to the
black hole, as has also been pointed out by e.g. Levinson \& Eichler (1993).

Some baryons are expected to be driven into the tower (along field lines 
attaching to the disk) by neutrino ablation (e.g., Levinson~2006).  
As of yet, no detailed calculations of neutrino-driven ablation exist 
for collapsar disks. For example, neutrinos may escape the disk in hot 
magnetically dominated bubbles without advecting large numbers of baryons 
into the tower.
In addition, Vlahakis~et~al. (2003) argued that neutrons, which may 
constitute the majority of the baryons in the disk wind, are not subject 
to electromagnetic acceleration and hence decouple from the relativistic 
outflow at rather moderate Lorentz factors; this significantly lessens 
the baryon pollution expected from the disk wind.

%****************************************************************

\subsection{Effect of MHD Instabilities on the Tower Evolution}
\label{subsec-stability}

One of the most serious uncertainties in the magnetic tower model
is the stability of this highly twisted structure. In this section 
we discuss the effects that ideal MHD instabilities may have on the 
magnetic tower in the collapsar context. A scenario involving 
reconnection inside the tower due to finite resistivity will be 
discussed in the next section.

Indeed, as the tower grows, it becomes very elongated and hence prone 
to a number of ideal MHD instabilities. Among the instability candidates 
that first come to mind there are: 

({\it i}) Rayleigh--Taylor (or, perhaps more relevant, 
Richtmyer--Meshkov) instability taking place at the 
tower's top. This may cause splitting of the magnetic 
tower into a number of separate strands interlaced with 
regions filled with stellar matter.

({\it ii}) Kelvin-Helmholtz instability at the side boundary 
between the tower and the cocoon, leading to entrainment of 
the baryonic stellar material into the tower. This instability 
however should be stabilized by the strong vertical magnetic 
field in the tower.

({\it iii}) A non-axisymmetric ideal-MHD kink-like instability taking place
in the main body of the tower. Although this problem has not been studied in
the magnetic-tower context, it has been considered in the framework of
relativistic force-free electrodynamics for field lines connecting a disk 
to a black hole (Gruzinov 1999). In addition, an important clue can be 
derived from stability studies of regular MHD jets. For example, a recent 
numerical investigation by Nakamura \& Meier (2004) indicated that, when 
a non-relativistic Poynting-flux dominated jet propagates through a 
stratified external medium, the jet stability strongly depends on 
the background density and pressure profiles along the jet. In particular, 
these authors found that a steep external pressure gradient forestalls
the instability onset.

If it does go unstable, the kink is probably the most dangerous instability
that may lead, in the nonlinear phase, to a significant, although perhaps
temporary, disruption.  Such a disruption, however, is not necessarily a bad
thing: the tower may be able to reform after being disrupted (as is seen in
laboratory experiments by Lebedev~et~al. 2005) and the resulting non-steady
evolution may provide a plausible mechanism for rapid variability seen in
gamma-ray bursts.

We also would like to point out that, as one expected outcome of the
nonlinear development of an MHD instability, a significant fraction of the
toroidal magnetic field energy may be dissipated into thermal energy
(Eichler~1993; Begelman~1998).  Importantly, this process may occur somewhere
inside the tower, way above the inner accretion disk. Since the plasma
density in the tower is very low, this may have an additional beneficial
side-effect on the explosion dynamics due to the possibility of non-thermal
neutrino emission coming from the magnetic dissipation sites. The resulting
high-energy neutrinos are expected to have better coupling to baryonic
stellar matter, thus providing an extra source for the strong shock
propagating through the star (Ramirez-Ruiz \& Socrates~2005).  In addition,
as has been shown by Drenkhahn \& Spruit (2002; see also Giannios \& Spruit
2006), magnetic energy dissipation may also have a positive influence on the
acceleration of the Poynting-flux dominated outflow.

So far as we know, there have been no formal stability studies 
of magnetic towers to date. Such studies are clearly needed. 
They may involve a linear perturbation analysis or a 
non-axisymmetric MHD simulation. Such stability studies 
should take into account several stabilizing effects. 
First, Tomimatsu et al. (2001) have conducted a linear 
stability analysis of the kink instability in a narrow 
rotating relativistic force-free jet and have shown that 
rapid field-line rotation inhibits the instability. A second 
important stabilizing factor, discussed in~\S~\ref{subsec-rel}, 
is the expectation that the tower should quickly transition to the 
relativistic regime. Once the outflow becomes ultra-relativistic 
with a very large $\gamma$-factor, the relativistic time delay may 
effectively stabilize the tower (Giannios \& Spruit 2006). The reason 
for this is that MHD instabilities grow on the local Alfv\'en-crossing 
time in the fluid frame and hence much slower in the laboratory frame. 
As a result, even if instabilities are excited, they do not have enough 
time to develop before the break-out of the flow from the star. 
Finally, the magnetic tower is not in a vacuum, but is being confined 
by a high-pressure cocoon. The cocoon may provide some stabilization
at least for external kink modes, although internal kink modes will 
not be affected. Some theoretical evidence supporting the idea of
the external pressure stabilization can be derived from K\"onigl 
\& Choudhuri's (1985) analysis of a force-free magnetized jet that
confined by an external pressure. Indeed, they showed that a 
non-axisymmetric helical equilibrium state becomes energetically 
favorable (conserving the total magnetic helicity in the jet) 
only when the pressure drops below a certain critical value. 
If this happens and the external kink does go unstable, then
this non-axisymmetric equilibrium may perhaps be interpreted 
as the end point of the non-linear development of the instability.

%****************************************************************

\subsection{Reconection Inside the Tower}
\label{subsec-reconnection}

Another important process that may affect the propagation of the magnetic tower
through the star is magnetic reconnection across the cylindrical separatrix 
current sheet at~$R=R_s$. This process could, in principle, lead to the break 
up of a single tower into a train of smaller spheromak-like plasmoids. 
Instead of a further lengthening of the tower, one would then effectively 
get continuous injection of new plasmoids at the base of the outflow. 
The new plasmoids would push the existing ones further up the axis, 
so that the net dynamical effect would probably be essentially the 
same as that of one single continuously growing tower. The physics of 
this process is in fact identical to the cyclic generation of plasmoids 
via reconnection across the separatrix suggested in the context of 
magnetospheres of accreting young stars, resulting in a knotty jet 
(Goodson~et~al. 1999; see also Uzdensky~2004). At what level this 
tearing instability should saturate, i.e., what should be the expected 
size of the plasmoids, is presently not known.

We would like to remark, however, that we doubt that a fast and 
efficient reconnection process is possible in the tower, as long 
as it is still deep inside the star. It is indeed true that fast 
reconnection is commonly believed to be the underlying mechanism 
for many spectacular and most energetic astrophysical phenomena, 
from substorms the Earth magnetosphere, to flares in the solar 
corona (e.g., Sweet~1958), to X-ray flares in magnetars (Lyutikov~2003). 
However, all these cases are characterized by relatively low-density 
environments, where classical collisional resistivity is so small 
that other, non-classical terms in the generalized Ohm law (i.e., 
the Hall term or anomalous resistivity) become important. 
This happens when the reconnection layer thickness, computed from 
the classical Sweet--Parker theory (Sweet~1958; Parker~1957) with 
Spitzer resistivity, becomes smaller than the relevant microscopic 
plasma length scales, such as the collisionless ion skin depth and/or 
the ion gyro-radius. This condition is easily satisfied both in the solar 
corona and in the Earth magnetosphere. Then, as many numerical simulations 
(e.g., Mandt~et~al. 1994) have shown, reconnection proceeds at a fast rate, 
in part due to the so-called Petschek mechanism (Petschek~1964).

However, in the dense collapsar environment the situation is different.
Even if the plasma pressure inside the magnetic tower constitutes a tiny
but finite fraction  of the magnetic pressure, the resulting electron-positron 
pair density is so high that the plasma should be considered strongly 
collisional. For example, let us assume, as an illustration, that the
interior of the tower is filled with an optically-thick radiation-pressure
dominated pair plasma with the thermal energy density equal to~$10^{-3}$
of the magnetic energy density and let us make some simple estimates 
relevant to reconnection physics. For a characteristic magnetic field 
strength of~$10^{14}~{\rm G}$, the corresponding plasma temperature 
should be of order $T\sim 3\cdot 10^9~{\rm K}\simeq 300~{\rm keV}$. 
The corresponding equilibrium pair density [using the non-relativistic 
formula $n_e=2(m_e k_B T/2\pi\hbar^2)^{3/2}\,\exp(-m_ec^2/k_B T)$ for 
a rough estimate] is on the order of~$n_e\sim 2\cdot 10^{29}~{\rm cm}^{-3}$.
The classical non-relativistic Spitzer resistivity due to electron-positron
collisions at the above temperature is~$\eta\sim 0.1~{\rm cm}^2/{\rm sec}$.
Now, the main dimensionless quantity that characterizes the reconnection
layer in resistive MHD regime is the Lundquist number $S\equiv V_A L/\eta$,
where~$L$ is the global length-scale of the reconnecting system.
Taking $L=10^7~{\rm cm}$ and using the speed of light instead of 
the Alfv\'en velocity, we get $S\sim 3\cdot 10^{18}$. Then, the 
classical Sweet--Parker reconnection theory gives us an estimate 
for the thickness of the reconnection layer as small as 
$\delta_{\rm SP}=LS^{-1/2}\sim 6\cdot 10^{-3}~{\rm cm}$.
This thickness is of course tiny compared with~$L$, but 
it is huge compared with the relevant physical plasma 
length-scales, such as the electron collisional mean 
free path ($l_{\rm mfp}\sim 10^{-6}~{\rm cm}$),
the electron gyro-radius ($\rho_e\sim 10^{-11}~{\rm cm}$),
and the electron collisionless skin-depth 
($d_e\equiv c/\omega_{pe}\sim 10^{-9}~{\rm cm}$).
This means that, although they are themselves very small, 
the classical collisional resistivity (due to $e^+-e^-$ 
Coulomb collisions) and the Compton drag (due to $e^+-\gamma$ 
and $e^--\gamma$ collisions) greatly exceed the collisionless 
terms in the generalized Ohm law. Therefore, resistive MHD 
(augmented by the Compton drag which may give a comparable, 
perhaps even somewhat larger than Spitzer, contribution to 
the total resistivity) should provide an accurate description 
of the plasma, even inside a very thin reconnection layer. 

During the past two decades it has been convincingly demonstrated, 
via numerical simulations (e.g., Biskamp~1986; Uzdensky \& Kulsrud~2000), 
theoretical analysis (Kulsrud~2001; Malyshkin~et~al. 2005), and laboratory 
plasma experiments (Ji~et~al.~1999), that in collisional resistive-MHD 
systems reconnection proceeds in the very slow Sweet--Parker regime and 
that the Petschek fast reconnection mechanism is disabled. 
If we can extend this result to the highly relativistic and 
optically-thick electron-positron plasma inside the collapsar 
magnetic tower, we then have to conclude that any large-scale 
reconnection across the separatrix is very slow and inefficient. 
One has to be aware of several caveats, however.
First, the role of the photon drag on reconnection dynamics is not 
known, especially its effect on the geometry of reconnecting flow. 
Second, there is a possibility that, even in the resistive-MHD regime, 
reconnection can be greatly enhanced in the presence of MHD turbulence, 
as suggested by Lazarian \& Vishniac (1999). And third, since magnetic 
energy density in the tower greatly exceeds both the plasma pressure 
and its rest-mass energy density, reconnection flow is bound to be 
strongly relativistic; then special-relativistic effects, such as 
the Lorentz contraction, may modify the conclusions derived from 
the non-relativistic reconnection theory (Blackman \& Fields 1994;
Lyutikov \& Uzdensky 2003). With all these reservations in mind, 
we still believe that reconnection processes are strongly inhibited
in the tower when it is still deep inside the collapsing star.
However, as the tower grows and eventually breaks out of the star,
the plasma cools and the particle density in it drops rapidly. 
Therefore, we should expect that the plasma will become collisionless, 
at least as far as reconnection physics is concerned, at late times,
perhaps after the break-out. This opens up the possibility that any 
possible disruption of the tower through reconnection (and corresponding 
magnetic energy release) is delayed and happens only after the tower 
clears the stellar surface.

%****************************************************************

\subsection{Prospects for Numerical Simulations}
\label{subsec-numerical}

In this paper we have developed a mathematical framework describing a
sequence of static force-free equilibria which represents a magnetic tower
inside a star.  To verify if our picture is realizable in practice and to
make further progress in understanding the development and evolution of
magnetic towers in stars, it is desirable to perform time-dependent numerical
simulations.  Such simulations will determine the degree to which the
simplifications we have adopted for our analytic model hold in realistic
environments.  The key feature to be investigated is the structure of the
tower and the cocoon and how they may be affected by instabilities.

We envision a sequence a numerical simulations in which increasingly
realistic physics is incrementally added to the simplest elemental
simulation.  Each of these studies will be able to answer a subset of key
questions with increasing degree of realism.  First, the following questions
can be addressed with non-relativistic axisymmetric MHD simulations.  What
basic magnetic field configuration results, in practice, when the conditions
we describe are set up?  Under what conditions does a magnetic tower form, if
at all?  How do the tower growth and structure change as the tower expands
into regions of lower density?  To what degree is cocoon material mixed into
the tower, and magnetic field from the tower mixed into the cocoon?  How
strongly is magnetic field concentrated toward the axis?  How does the
Poynting flux depend on radius and height?  What is the overall field line
geometry?  Is magneto-centrifugal acceleration from the base of the outflow
inhibited?  In simulations where multiple towers are investigated, how do
they interact?  To what degree does the cocoon help collimate and stabilize
the tower? How rapidly do the cocoon walls spread laterally?  Can the cocoon
expansion result in the disruption of the star?  What differences are there
between cocoons formed by purely hydrodynamical jets and by magnetic towers?

Next, relativistic MHD simulations will be able to address fundamental
questions of critical interest to the application of magnetic towers to
gamma-ray bursts.  Of particular interest is the beaming angle and angular
distribution of energy, Lorentz factor and magnetic field after the tower
makes the transition to relativistic expansion and when it eventually breaks
out of the stellar surface.  Do relativistic effects suppress the growth of
instabilities in the tower and surrounding plasma?  In addition, inclusion of
relativity is necessary for some regions of the central engine, e.g. outside
of a magnetar light cylinder or the inner parts of black hole accretion
disks.

Finally, three-dimensional simulations, both relativistic and non-relativistic,
will enable study of non-axisymmetric instabilities.  Intermixing of baryonic
material and magnetic field between the tower and the cocoon may also be
quantitatively addressed with three-dimensional simulations.

Several numerical schemes have recently been developed (Koide, Shibata \&
Kudoh 1999; Gammie, McKinney \& T\'oth 2003; Del Zanna, Bucciantini \&
Londrillo 2003; De Villiers, Hawley \& Krolik 2003; Fragile 2005; Komissarov
2005; Nishikawa et al. 2005) which should be capable of simulating magnetic
towers powered by central engines formed after core collapse. In fact,
general relativistic MHD simulations of collapsars have already been
performed by two groups (Mizuno et al. 2004; De Villiers, Staff, \& Ouyed
2005).  In principle these simulations may contain structures which could be
termed magnetic towers as described in this paper.  However, it may be
difficult to clearly discern the properties of the magnetic structure in
complex simulations.  For this reason we suggest above incremental
simulations guided by analytic studies as presented here.

While the magnetic tower itself is force-free, pressure and inertia of the
surrounding plasma play a key role in its development.  This points to the
desirability of magnetohydrodynamical (MHD) simulations capable of
simultaneously solving force-free and hydrodynamical regions of the flow.
One possible choice is a hybrid numerical scheme which solves relativistic
force-free equations in highly magnetized parts of the computational domain,
e.g. within the tower, with a moving contact discontinuity evolved within a
hydrodynamical simulation (Blandford 2005, private communication).
Another possible strategy is to evolve relativistic MHD equations, 
but switch to relativistic force-free equations whenever $B^2$ 
becomes larger than~$\rho c^2$ (e.g., McKinney~2006).

As an ultimate goal, numerical simulations of the full problem, including a
detailed description of the central engine with relevant microphysical
processes and neutrino transport, are desirable for a comprehensive
understanding of the formation and evolution of a magnetic tower in a star.

%*********************************************************************

\section{Conclusions}
\label{sec-conclusions}

In this paper we have proposed a new mechanism for driving 
a strongly-collimated, baryon-poor, magnetically-dominated
outflow through a massive star in the collapsar model (e.g.,
MacFadyen \& Woosley 1999) for long-duration gamma-ray bursts. 
This magnetic mechanism may work independently and in parallel
with the usual neutrino-driven outflow mechanism. It may also 
be relevant for asymmetric core-collapse supernova explosions. 

To model the magnetically-dominated outflow, we invoke the concept 
of the magnetic tower, first introduced by Lynden-Bell (1996) in 
the AGN context. Overall, our model can be outlined as follows. 
The core collapse of a massive rotating star may result in two distinct
configurations, both plausible candidates for the GRB central engine.
The first one is a stellar mass black hole accreting rapidly through
an accretion disk. The second is a millisecond magnetar. A hybrid
configuration with a disk around a magnetar is also possible.  
In the first case, the core of a rotating massive star collapses into 
a black hole (or forms a proto-neutron star), whereas the overlying 
stellar material that is continuously falling towards the center may 
possess enough angular momentum to form an accretion disk. 
Any pre-existing stellar magnetic field becomes greatly amplified by 
the combined action of compression during the collapse and field-line 
stretching due to differential rotation. In addition, the disk is 
generally unstable to magneto-rotational instability (MRI) and becomes 
turbulent. This results in further amplification of the magnetic field 
due to turbulent dynamo action inside the disk, up to the level that may 
be in excess of~$10^{16}\ \gauss$ (as suggested by Akiyama~et~al. 2003 
in the core-collapse supernova context). Even though this field is
predominantly toroidal, the poloidal field component may also be
significant; we believe that a figure of $10^{15}\ \gauss$ for the
large-scale poloidal magnetic field is not unreasonable.

As a result of the amplification, magnetic tubes become buoyant 
and escape out of the disk into the overlying low-density corona, 
forming coronal loops. These loops are anchored at both ends 
at the disk surface but, generally speaking, the two footpoints 
of a loop are at different distances from the center and hence
rotate at different angular velocities. The resulting differential 
rotation further twists the magnetic loops, generating more toroidal 
flux, and the disk magnetosphere tends to inflate. A similar process 
has also been shown to work in the case of a magnetic field linking
the disk directly to the black hole (Uzdensky~2005). In any case, it 
is this inflation process that is ultimately responsible for the formation 
of the magnetically-dominated outflow in our model. For simplicity, 
we consider the evolution of an axisymmetric magnetic arcade. 
Because the density in the corona above the disk is low,
the initial expansion is force-free and takes place at some finite 
angle with respect to the rotation axis. Subsequently, however, 
the expanding magnetic field weakens to such a degree that it starts
to feel the presence of the ambient stellar gas. As was shown by Lynden-Bell
(1996), any external gas pressure surrounding an expanding twisted 
magnetic arcade eventually becomes dynamically important and channels
the expansion into the vertical direction; a tall magnetic tower forms
and continuously grows as a result. In our model, we modify Lynden-Bell's
model by taking into account that the tower expansion is supersonic
with respect to the unperturbed stellar gas. We thus envision the 
growing magnetic tower acting as a piston that drives a strong shock 
through the star. The hot shocked stellar material between the shock
and the tower forms a high-pressure cocoon, similar to that seen in
hydrodynamic simulations by Zhang~et~al. (2003). This cocoon envelopes
the tower and provides the collimating pressure for it. Thus, in our 
model, the tower is confined not by the pressure of the background 
stellar material, but by its inertia; the strong shock and the cocoon
act as mediators that convert the inertial support into the pressure 
support that ultimately acts on the tower.

The entire configuration grows vertically with time and eventually reaches
the star's surface, thus providing a very narrow baryon-clean channel in the
form of a Poynting-flux dominated jet, surrounded by a less-collimated hot
gas outflow. One of the great advantages of the magnetic tower model in the
GRB context is that the magnetic field lines remain closed (e.g., with
both ends tied to the disk) during the expansion process and, as a result,
the volume occupied by the tower remains insulated from the surrounding
stellar gas and hence relatively baryon-free.  As the tower grows, it just
pushes the stellar gas out of its way into the cocoon.

In addition, we believe that the physical scenario developed in this paper
may also be applicable to the case where the central engine is a
rapidly-rotating (i.e., millisecond) magnetar. The rotational energy of the
magnetar is extracted magnetically as a Poynting flux that inflates a
baryon-free cavity inside the progenitor star. This mechanism is similar to
that proposed by Ostriker \& Gunn (1971) for powering type~II supernovae.
However, whereas their model was spherically-symmetric, we suggest that a
relativistic generalization of the magnetic tower model can naturally result
in splitting of the magnetically-dominated outflow into a pair of oppositely
directed narrow jets.

In our paper we first give a simple description of Lynden-Bell's magnetic
tower model in its original accretion disk context
(\S~\ref{subsec-LB-model}). Then we present our model of a magnetic tower
propagating and driving a shock through a star
(\S~\ref{subsec-cocoon-shock}). We derive some basic scalings for the tower
parameters (\S~\ref{subsec-estimates}) and make some simple estimates for
them (\S~\ref{subsec-numbers}).  Then,
in~\S~\ref{sec-equations}--\ref{sec-examples}, we illustrate our model by two
specific analytical solutions for the magnetic field structure in the main
body of the tower; one of the solutions (\S~\ref{subsec-case-2}) is
characterized by a very narrow axial jet with a non-zero poloidal line
current present in the core of the tower.  In \S~\ref{sec-discussion} we
briefly discuss several interesting physical issues that, although important,
lie beyond the scope of the present paper. Namely, in \S~\ref{subsec-rel} we
argue that, even though it may start off non-relativistically, the expansion
of the tower will quickly have to transition to the relativistic regime.
In~\S~\ref{subsec-baryons} we discuss the issue of baryon contamination of
the tower by disk winds. Next, in~\S~\ref{subsec-stability} we touch upon the
question of stability of the magnetic tower configuration and address the
role that ideal MHD instabilities may play in our scenario.  In
\S~\ref{subsec-reconnection} we discuss the possibility of the tower
disruption and its break-up into a series of smaller plasmoids via internal
reconnection. We argue, however, that, while the tower is still deep inside
the star, the plasma inside of it is likely to be strongly collisional, so
that classical resistive-MHD should apply inside the reconnection region.  If
this is the case, then, we believe, any reconnection process should be very
slow, and so a global disruption through reconnection is not likely during
this stage. During subsequent expansion, however, the density drops and the
reconnection process may enter a faster collisionless regime, perhaps
resulting in rapid magnetic energy dissipation.

Finally, in accordance with our discussion in \S~\ref{sec-discussion}, 
we can formulate the following three areas that, we believe, require 
immediate attention and should and can be addressed in the near future.
They are: (1) a special-relativistic generalization of the magnetic 
tower model; (2) 3D relativistic MHD numerical simulations of a magnetic 
tower inside a star; (3) the conditions for and the non-linear development 
of, various MHD instabilities.

It is our pleasure to thank E.~Blackman, A.~Gruzinov, R.~Kulsrud, 
J.~McKinney, D.~Meier, A.~Spitkovsky, H.~Spruit, and C.~Thompson 
for stimulating discussions and comments.  
We are grateful to the Kavli Institute for Theoretical Physics 
at UC Santa Barbara for its hospitality during the program on 
the Physics of Astrophysical Outflows and Accretion Disks in 
Spring 2005 where part of this research was conducted.  
AIM acknowledges support from the Keck Fellowship at the Institute 
for Advanced Study. DAU's research has been supported by the National 
Science Foundation under Grant~PHY-0215581 (PFC: Center for Magnetic 
Self-Organization in Laboratory and Astrophysical Plasmas).

%****************************************************************

\section*{REFERENCES}
\parindent 0 pt

Akiyama, S., Wheeler, J.~C., Meier, D.~L., \& Lichtenstadt, I. 2003,
ApJ, 584, 954

Ardeljan, N.~V., Bisnovatyi-Kogan, G.~S., \& Moiseenko, S.~G. 2005, 
MNRAS, 359, 333

Begelman, M.~C. 1998, ApJ, 493, 291

Begelman, M.~C., Blandford, R and Rees, M,, Review of Modern Physics 1984,
56, 255

Biskamp, D. 1986, Phys. Fluids, 29, 1520

Blackman, E.~G. \& Field, G.~B. 1994, Phys. Rev. Lett., 72, 494

Blackman, E.~G., Nordhaus, J.~T., \& Thomas, J.~H. 2006, New Astronomy, 11, 452

Blandford, R.~D., \& Payne, D.~G.\ 1982, MNRAS, 199, 883 
 
Blandford, R.~D. \& Znajek, R.~L. 1977, MNRAS, 179, 433

Braithwaite, J., \& Spruit, H.~C.\ 2004, Nature, 431, 819 

Del Zanna, L., Bucciantini, N., \& Londrillo, P. 2003, A\&A, 400, 397	

De Villiers, J.-P., Hawley, J.~F., \& Krolik, J.~H.\ 2003, ApJ, 599, 1238 

De Villiers, J.-P., Staff, J., \& Ouyed, R. 2005; preprint (astro-ph/0502225)

Drenkhahn, G., \& Spruit, H. 2002, A\&A, 391, 1141

Eichler, D. 1993, ApJ, 419, 111

Fragile, P. C., 2005, preprint (astro-ph/0503305)

Fryer, C.~L., \& Heger, A. 2005, ApJ, 623, 302 

Fukuda, I. 1982, PASP, 94, 271 

Galama, T.~J. et al. 1998, Nature, 395, 670

Gammie, C. F., McKinney, J. C., \& T\'oth, G. 2003, ApJ, 589, 444

Giannios, D., \& Spruit, H. 2006, A\&A, 450, 887

Goodman, J., Dar, A., \& Nussinov, S. 1987, ApJ, 314, L7

Goodson, A.~P., B\"ohm, K.-H., \& Winglee, R.~M. 1999, ApJ, 524, 142

Gruzinov, A., 1999, preprint (astro-ph/9908101)

Harrison, F.~A. et al. 1999, ApJ, 523, L121

Heger, A., Woosley, S.~E., \& Spruit, H.~C. 2005, ApJ, 626, 350 

Hjorth, J. et al. 2003, Nature, 423, 847

Hawley, J.~F. \& Krolik, J.~H. 2006, ApJ, 641, 103

Hsu, S. \& Bellan, P.~M. 2002, MNRAS, 334, 257

%Lebedev, S.~V., Ciardi, A., Ampleford, D.~J., Bland, S.~N., 
%Bott, S.~C., Chittenden, J.~P., Hall, G.~N., Rapley, J., 
%Jennings, C.~A., Frank, A., Blackman, E.~G., \& Lery, T.
%2005, MNRAS, 361, 97

Lebedev, S.~V. et al. 2005, MNRAS, 361, 97

Hirschi, R., Meynet, G., \& Maeder, A. 2005, A\&A, 443, 581 

Ji, H., Yamada, M., Hsu, S., Kulsrud, R., Carter, T., \& Zaharia, S. 1999,
Phys. Plasmas, 6, 1743

Kato, Y., Hayashi, M.~R., \& Matsumoto, R. 2004, ApJ, 600, 338

Koide, S., Shibata, K., \& Kudoh, T. 1999, ApJ, 522, 727

Komissarov, S. S. 2005, MNRAS, 359, 801

K{\"o}nigl, A. \& Choudhuri, A.~R. 1985, ApJ, 289, 173

Kulsrud, R.~M. 2001, Earth, Planets and Space, 53, 417

Kulsrud, R.~M. 2005, Plasma Physics for Astrophysics,
(Princeton: Princeton Univ. Press)

Lazarian, A. \& Vishniac, E.~T. 1999, ApJ, 517, 700

LeBlanc, J.~M. \& Wilson, J.~R. 1970, ApJ, 161, 541

Lee, H.~K., Wijers, R.~A.~M.~J., \& Brown, G.~E. 2000, Phys. Reports, 325, 83

Lei, W.-H., Wang, D.-X., \& Ma, R.-Y. 2005; ChJAA, 5, Suppl., 279

Levinson, A. \& Eichler, D. 1993, ApJ, 418, 386

Levinson, A. 2006; preprint (astro-ph/0602358)

Li, H., Lovelace, R.~V.~E., Finn, J.~M., \& Colgate, S.~A. 2001, ApJ, 561, 915 

Lovelace, R.~V.~E., Romanova, M.~M., \& Bisnovatyi-Kogan, G.~S. 1995, 
MNRAS, 275, 244

Lovelace, R.~V.~E. \& Romanova, M.~M. 2003, ApJ, 596, L159

Lynden-Bell, D. \& Boily, C. 1994, MNRAS, 267, 146

Lynden-Bell, D. 1996, MNRAS, 279, 389

Lynden-Bell, D. 2003, MNRAS, 341, 1360

Lyutikov, M. \& Uzdensky, D. 2003, ApJ, 589, 893

Lyutikov, M. \& Blandford, R. 2003; preprint (astro-ph/0312347)

Lyutikov, M. 2004; preprint (astro-ph/0409489)

Lyutikov, M. 2006 to appear in the New Journal of Physics; 
preprint (astro-ph/0512342)

MacFadyen, A.~I. \& Woosley, S.~E. 1999, ApJ, 524, 262

MacFadyen, A.~I., Woosley, S.~E., \& Heger, A.\ 2001, ApJ, 550, 410 

Maeder, A., \& Meynet, G. 2005, A\&A, 440, 1041 

Malesani, D., et al. 2004, ApJL, 609, L5 

Malyshkin, L.~M., Linde,T., \& Kulsrud, R.~M. 2005,
Phys. Plasmas, 12, 102902

%Mandt, M.~E., Denton, R.~E., \& Drake, J.~F. 1994,
%Geophys. Res. Lett., 21, 73

Mandt, M.~E. et al. 1994, Geophys. Res. Lett., 21, 73

Matheson, T., et al. 2003, ApJ, 599, 394 

Matzner, C.~D. 2003, MNRAS, 345, 575

McKinney, J.~C. 2005, ApJ, 630, L5

McKinney, J.~C. 2006, accepted to MNRAS; preprint (astro-ph/0601410)

Meier, D.~L., Epstein, R.~I., Arnett, W.~D., \& Schramm, D.~N. 1976,
ApJ, 204, 869

Meszaros, P. \& Rees, M.~J. 1997, ApJ, 482, L29 

Mizuno, Y., Yamada, S., Koide, S., \& Shibata, K. 2004, ApJ, 615, 389 

Nakamura, M. \& Meier, D.~L 2004, ApJ, 617, 123

Nishikawa, K.-I., Richardson, G., Koide, S., Shibata, K., Kudoh, T., Hardee,
P., \& Fishman, G.~J.\ 2005, ApJ, 625, 60

Ostriker, J.~P. \& Gunn, J.~E. 1971, ApJ, 164, L95

Paczynski, B. 1998, ApJ 494, L45

Parker, E.~N. 1957, J. Geophys. Res., 62, 509

Petschek, H.~E. 1964, AAS-NASA Symposium on Solar Flares, 
(National Aeronautics and Space Administration, Washington, 
DC, 1964), NASA SP50, 425.

Petrovic, J., Langer, N., Yoon, S.-C., \& Heger, A. 2005, A\&A, 435, 247 

Proga, D., MacFadyen, A.~I., Armitage, P.~J., \& Begelman, M.~C. 2003, 
ApJ, 599, L5

Proga, D. \& Zhang, B. 2006, submitted to ApJ Letters; 
preprint (astro-ph/0601272)

Qian, Y.-Z., \& Woosley, S.~E.\ 1996, ApJ, 471, 331 

Ramirez-Ruiz, E., Celotti, A., \& Rees, M.~J. 2002, MNRAS, 337, 1349

Ramirez-Ruiz, E., \& Socrates, A. 2005, preprint (astro-ph/0504257)

Soderberg, A.~M., et  al. 2005, ApJ, 627, 877 

Soderberg, A.~M., et al. 2006a, ApJ, 636, 391 

Soderberg, A.~M., Nakar, U., Berger E., \& Kulkarni S., 2006b, ApJ, 638, 930

Stanek, K.~Z., Garnavich, P.~M., Kaluzny, J., Pych, W., \& Thompson, I.
1999, ApJ, 522, L39

Stanek, K.~Z. et al. 2003, ApJ, 591, L17

Sweet, P.~A. 1958, in Electromagnetic Phenomena in Cosmical Physics, 
ed.~B.~Lehnert, (New York: Cambridge Univ. Press), 123

Thompson, C.\ 1994, MNRAS, 270, 480 

Thompson, T.~A., Chang, P., \& Quataert, E.\ 2004, ApJ, 611, 380 

Tomimatsu, A., Matsuoka, T., \& Takahashi, M. 2001, Phys. Rev. D.,
64, 123003

Usov, V.~V.\ 1992, Nature, 357, 472 
 
Uzdensky, D.~A. \& Kulsrud, R.~M. 2000, Phys. Plasmas, 7, 4018

Uzdensky, D.~A., K\"{o}nigl, A., \& Litwin, C. 2002, ApJ, 565, 1191

Uzdensky, D.~A., 2002, ApJ, 574, 1011

Uzdensky, D.~A. 2004, Ap\&SS, 292, 573

Uzdensky, D.~A. 2005, ApJ, 620, 889

van Putten, M.~H.~P.~M. \& Ostriker, E.~C. 2001, ApJ, 552, L31

van Putten, M.~H.~P.~M. \& Levinson, A. 2003, ApJ, 584, 937

Vlahakis, N. \& K{\"o}nigl, A. 2001, ApJ, 563, L129

Vlahakis, N., Peng, F., \& K{\"o}nigl, A. 2003, ApJ, 594, L23

Wheeler, J.~C., Yi, I., H\"oflich, P., \& Wang, L. 2000, ApJ, 537, 810

Wheeler, J.~C., Meier, D.~L., \& Wilson, J.~R. 2002, ApJ, 568, 807

Woosley, S.~E. 1993, ApJ, 405, 273 

%Woosley, S.~E. \& Bloom, J. 2006, ARAA, in press

Yi, I. \& Blackman, E.~G. 1998, ApJ, 494, L163

Yoon, S.-C., \& Langer, N. 2005, A\&A, 443, 643 

Zeh, A., Klose, S., \& Hartmann, D.~H.\ 2004, ApJ, 609, 952 

Zhang, W., Woosley, S.~E., \& MacFadyen, A.~I. 2003 ApJ, 586, 356

%*****************************************************************

\end{document}